\documentclass[aps,prb,twocolumn,superscriptaddress,preprintnumbers]{revtex4-2}

\usepackage[T1]{fontenc}
\usepackage{amssymb}
\usepackage{graphicx}
\usepackage{color}
\usepackage{caption}
\usepackage{hyperref}
\usepackage{amsthm}
\usepackage{amsmath}
\usepackage{braket}
\usepackage{appendix}
\usepackage{subcaption}
\usepackage{comment}

\usepackage{tikz}
\usetikzlibrary{patterns}
\usepackage{physics}

\usepackage{amsthm}
\usepackage{enumitem}
\usepackage{comment}

\begin{document}


\title{
Simulating continuous symmetry models with discrete ones 
}

\author{A. G. Catalano}
\affiliation{Institut Ru\dj er Bo\v{s}kovi\'c, Bijeni\v{c}ka cesta 54, 10000 Zagreb, Croatia}
\affiliation{Universit\'e de Strasbourg, 4 Rue Blaise Pascal, 67081 Strasbourg, France}

\author{D. Brtan}
\affiliation{SISSA and INFN, via Bonomea 265, 34136 Trieste, Italy}

\author{F. Franchini}
\email{fabio@irb.hr}
\affiliation{Institut Ru\dj er Bo\v{s}kovi\'c, Bijeni\v{c}ka cesta 54, 10000 Zagreb, Croatia}

\author{S. M. Giampaolo}
\email{sgiampa@irb.hr}
\affiliation{Institut Ru\dj er Bo\v{s}kovi\'c, Bijeni\v{c}ka cesta 54, 10000 Zagreb, Croatia}


\preprint{RBI-ThPhys-2022-2}


\begin{abstract}
Especially in one dimension, models with discrete and continuous symmetries display different physical properties, starting from the existence of long-range order. In this work, we that, by adding topological frustration, an antiferromagnetic $XYZ$ spin chain, characterized by a discrete local symmetry, develops a region in parameter space which mimics the features of models with continuous symmetries. For instance, frustration closes the mass gap and we describe a continuous crossover between ground states with different quantum numbers, a finite (Fermi) momentum for low energy states, and the disappearance of the finite order parameter. Moreover, we observe non-trivial ground state degeneracies, non-vanishing chirality and a singular foliation of the ground state fidelity .
Across the boundary between this chiral region and the rest of the phase diagram any discontinuity in the energy derivatives vanishes in the thermodynamic limit.
\end{abstract}

\maketitle

\section{Introduction}

One-dimensional spin-1/2 systems have always represented a fundamental field of study for the physics of quantum many-body systems but, {quite surprisingly}, nowadays their relevance is {still} growing.
{Indeed, while in the past they have represented toy models to detect features to embed into frameworks of more realistic systems, recently, the development of novel experimental devices~\cite{Jaksch1998, Porras2004, Zippilli2014, Labuh2016} have disclosed the path toward direct tests of theoretical predictions.}
Assuming invariance under spatial translation and limiting to systems with finite range interactions, in agreement with Goldstone's theorem~\cite{Goldstone1962, Wreszinski1976}, they are usually classified into two large families. 
 
{The first includes models in which Hamiltonians show local continuous symmetries. 
In these cases, {the systems} admit sets of quantum numbers, i.e. sets of distinct eigenvalues of operators commuting with the Hamiltonian, whose size scales with the chain length. 
In these models, properly ordered phases are absent~\cite{MerminWagner} {even at zero temperature, due to quantum fluctuations. Another remarkable property of these system is that, at criticality (that is, when the mass gap vanishes)  a parameter change can trigger continuous cross-overs between} non-degenerate ground states with different quantum numbers.}
{To fix the ideas let us }{consider the example of the $XXZ$ spin-$1/2$ chain.}{ 
Such a model holds a continuous $U(1)$ rotational symmetry along the $z$-axis and, due to such symmetry, the eigenstates of the Hamiltonian can be classified by their total magnetization in the $z$ direction. 
{In a region of small interaction along $z$ (in modulus) without local fields,} the ground state has a vanishing total magnetization, and the energy spectrum is gapless in the thermodynamic limit~\cite{Franchini17}.
Turning on a magnetic field along $z$ induces a finite magnetization in the ground state preserving the criticality of the system up to a critical value.  
Moreover, together with the nonvanishing magnetization, the ground state also acquires a non-zero momentum, {although not a macroscopic one. Usually, the ground state is static, in order to minimize the kinetic energy, but here we show that, in presence of topological frustration, the lowest energy state of a system can be just stationary. Moreover, since low energy excitations are also characterized by momenta close to that of the ground state, this constitutes a {\it Fermi momentum}.
Note that, contrary to what happens in higher-dimensions, 1D models with continuous symmetry posses a Fermi momentum  independently from the statistics of the microscopical degrees of freedom~\cite{Giamarchi}.}
Thus, by changing the external field, the system undergoes a series of ground state cross-overs between non-degenerate states with different values of the total magnetization, i.e. different quantum numbers, that are clearly mutually orthogonal. 
Hence, in the thermodynamic limit, systems like the \textit{XXZ} chain in this regime show an extreme case of orthogonality catastrophe~\cite{Anderson1967-1, Anderson1967-2}:} {ground states of arbitrarily similar Hamiltonian are not only orthogonal in the thermodynamic limit, but at any finite size any change in the Hamiltonian parameters continuously moves between states with zero overlap~\cite{Kwok2008}.}

{On the opposite side, the second family of one-dimensional models is made by systems whose Hamiltonians display} {only discrete local symmetries, so that its eigenstates are characterized just by finite sets of quantum numbers. 
The physical properties of this second family are different from those above.
In their gapless regimes, that is when the mass gap closes algebraically with the system size, one cannot define a Fermi momentum for low energy excitations, and in the gapped case, these systems can develop a finite order parameter as a reflection of a spontaneously broken (discrete) symmetry.
In fact, few nearly degenerate low-energy states, possessing different quantum numbers, are separated by a finite energy gap from the rest of the spectrum. The gap between these states closes exponentially with the system size~\cite{Franchini17,Giamarchi} and in the thermodynamic limit the system can select a superposition of them as a ground state that violates a Hamiltonian's symmetry.
Such state is thus characterized by a non-zero order parameter, i.e. a non-vanishing expectation value for an operator that should otherwise vanish due to symmetry.}


{In this paper, we will show that it is possible to simulate the physics of models with continuous symmetries using models whose Hamiltonians possess only discrete symmetries. 
The key ingredient will be the introduction of a frustration~\cite{Toulouse1977, Vannimenus77, Wolf03, Sadoc1999, Diep2013, Giampaolo2011, Marzolino13} of topological origin in the latter models. 
Topological frustration can be induced in a system with short-range antiferromagnetic interactions through a special set of boundary conditions, the so-called frustrated boundary conditions (FBC)~\cite{Cador, Laumann2012, Dong2016, Giampaolo2018, Maric2020_destroy, Maric2020_induced}. They are realized assuming periodic boundary conditions in a system made of an odd number of spins.
In this setting, as described in detail in the rest of this work, the ground-state is described by an excitation delocalized along the whole chain. As a consequence, the length of the chain becomes a relevant scale for the system and, despite being just boundary conditions, FBCs can actually affect the thermodynamic limit and thus the physics of the system.}

{We will show that moving in a region of phase space, these systems undergo repeated crossovers between exactly orthogonal states. 
Since the discrete local symmetries of these models separate states only into finite sets (of quantum numbers), frustration further uses the representation of spatial translation to differentiate between the different ground states. Namely, depending on the Hamiltonian parameters, the ground states acquire finite momenta.
As discussed in a series of recent works, unless higher symmetries are consider which constrain the ground state momenta to specific values, FBC prevent the formation of a finite order parameter in the thermodynamic limit~\cite{Maric2020_destroy,Maric2020_induced,Maric2021_fate}, again mimicking the behavior of continuous symmetry systems.}

{We will illustrate this phenomenology in the framework of the topologically frustrated short-range anisotropic $XYZ$} {chain, a prototypical model featuring a $\mathbb{Z}_2$ symmetry, realized by the parity of the magnetization.}
With the help of both numerical diagonalization and analytical evaluations, we will prove that it {presents a region of parameter space}, which we name {\it chiral region}, in which even at finite sizes several two-fold degenerate manifolds play, alternatively, the role of ground states.
{Such manifolds, whose elements belong to the same parity, are completely identified by two eigenstates with opposite quantum numbers for the momentum operator.
Increasing the size of the system, the number of possible eigenvalues of the momentum grows, hence increasing also the number of crossovers in the chiral region.}
Thereby, in the thermodynamic limit, this system will show a behavior mimicking the critical phase of models with local continuous symmetry. 
Moreover, these ground state momenta act as Fermi points, and low energy excitations lie close to them.
The chiral region is separated from the rest of the parameter space by a {line at which the degeneracy of the ground state changes. 
However, none of the derivatives of the energy show non-analytic behavior. 
On the other hand, consistently with the fact that this transition is induced by the particular choice of boundary conditions, all the parameters we have analyzed in the attempt to classify the chiral region vanish algebraically with the chain length. 
The latter two properties support the idea that we are looking at a boundary-BKT-like phase transition, {but we are not aware of such occurrence in the literature, in any model}.}

{To highlight this peculiar picture, in the beginning, we will focus on different observables, such as chirality and the ground-state momentum. Such observables are known, in some cases, to be different from zero in systems with continuous symmetry such as the Heisenberg chain~\cite{Subrahmanyam1994, Subrahmanyam, Poilblanc}, but have never previously been observed in systems with discrete symmetries.
Moreover, to have a more direct characterization of the ground states in this region, we focus our attention on the ground-state fidelity.
This is a quantity directly borrowed from quantum information theory and it allows to appreciate how a system reacts to a small change of the Hamiltonian parameters~\cite{Zanardi2006, Zanardi2007_1, Venuti2007, Zanardi2007_2, You2009, Kolodubretz2013, Bhattacharya2014}.}

The paper is organized as follows. 
In the second section, we present several numerical results, obtained with an exact diagonalization approach, that depicts the main features of the chiral region. 
{In the third section, we study both the total ground-state fidelity, which drops to zero in the chiral region and the reduced one, obtained by considering the overlap between the reduced density matrices obtained by tracing out from the ground state all degrees of freedom but those at two sites.
To better analyze how these features scale in the thermodynamic limit, we restrict to the $XY$ chain in which we can exploit the fact that the model can be exactly mapped to a free-fermionic system and hence can be solved analytically. 
By employing such analytical solution, in the last section we show that both the violation of the invariance under spatial translation and the chirality vanish in the thermodynamic limit and that, at the boundary of the chiral region, none of the energy derivatives show non-analyticities.}
In the end, we draw our conclusions.

\section{Anisotropic XYZ chain}
\begin{figure}[t]
\includegraphics[width=\columnwidth]{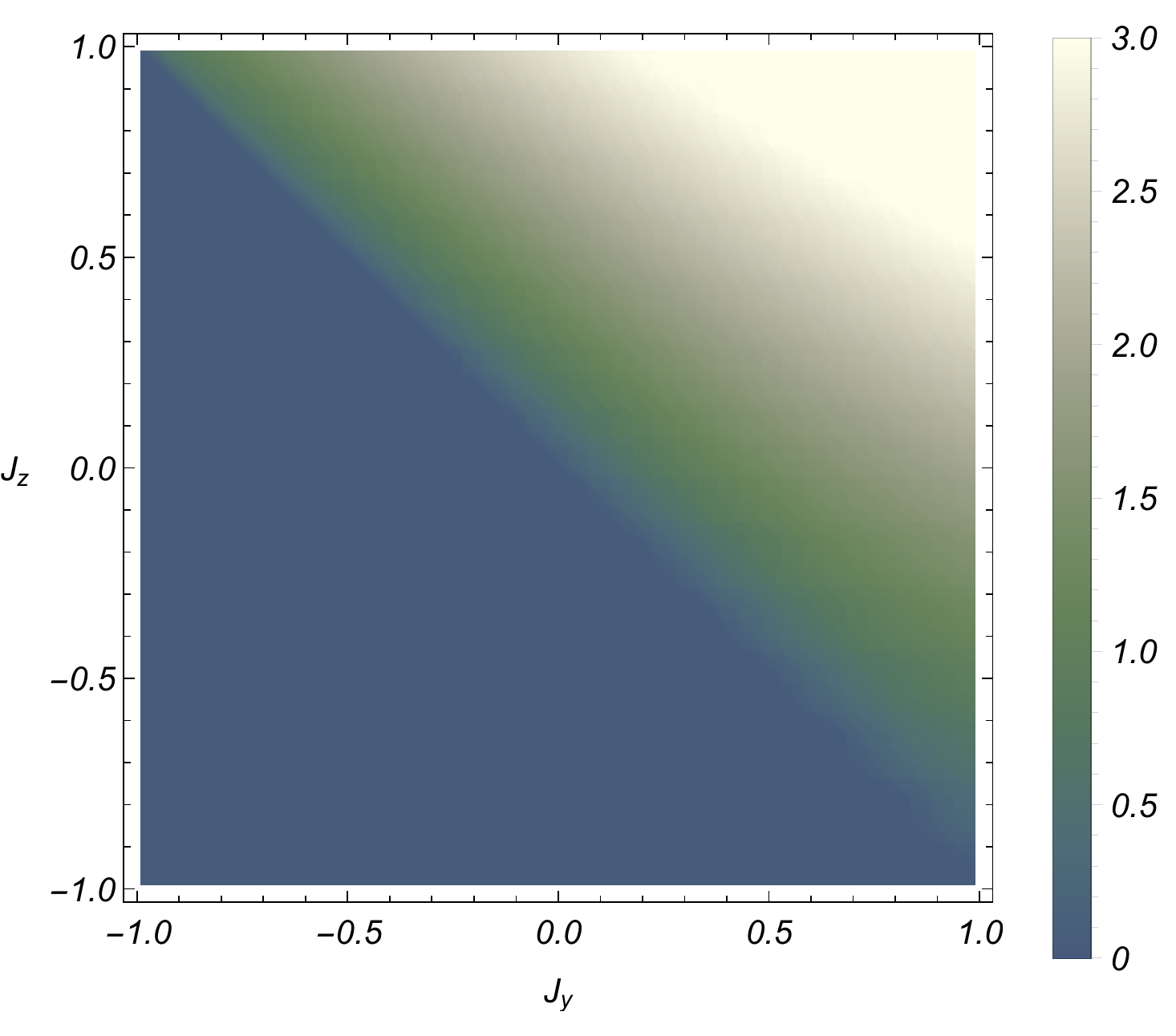}
\caption{Dependence of the threshold value $h^*$ as a function of $J_y\in(-1,1)$ and $J_z\in(-1,1)$ for $N=15$ and $J_x=1$. 
For $|h|<h^*$ (the {\it chiral region}) the ground state manifold is at least two-fold degenerate and spanned by states with finite and opposite momentum.}
\label{fig:criticalfield}
\end{figure}

{Let us start by introducing the model that we use in this paper as an example of the phenomenology we are presenting.}
The model is a very general short-range anisotropic Heisenberg chain with a local field that, without losing generality, we assume to align with the $z$ axis.
To induce topological frustration, we consider only the cases made of an odd number of sites with periodic boundary conditions in which the dominant interaction is antiferromagnetic. 
The Hamiltonian of such a system reads
\begin{equation}
\label{eq:Hamiltonian_ns}
    H=\sum_{j=1}^N \sum_{\alpha=x,y,z} J_\alpha \sigma^\alpha_{j}\sigma^\alpha_{j+1}
    -h\sum_{j=1}^N \sigma^z_{j},
\end{equation}
where $\sigma^\alpha_j$, for $\alpha=x,y,z$, are Pauli matrices and periodic boundary conditions require $\sigma^\alpha_j=\sigma^\alpha_{j+N}$. 
Moreover, we assume that $J_x\neq J_y$ to avoid that, the system acquires a continuous rotational symmetry around the $z$ axis, like the $XXZ$ chains shortly described in the introduction.  
As a consequence, the system holds only the discrete parity symmetry along $z$ $(\Pi^z=\prod_{i=1}^N\sigma^z_i, \left[\Pi^z,H\right]=0)$ that, independently of the system size, admits only two quantum numbers ($\pm1$).  

\begin{figure}[t]
\includegraphics[width=0.95\columnwidth]{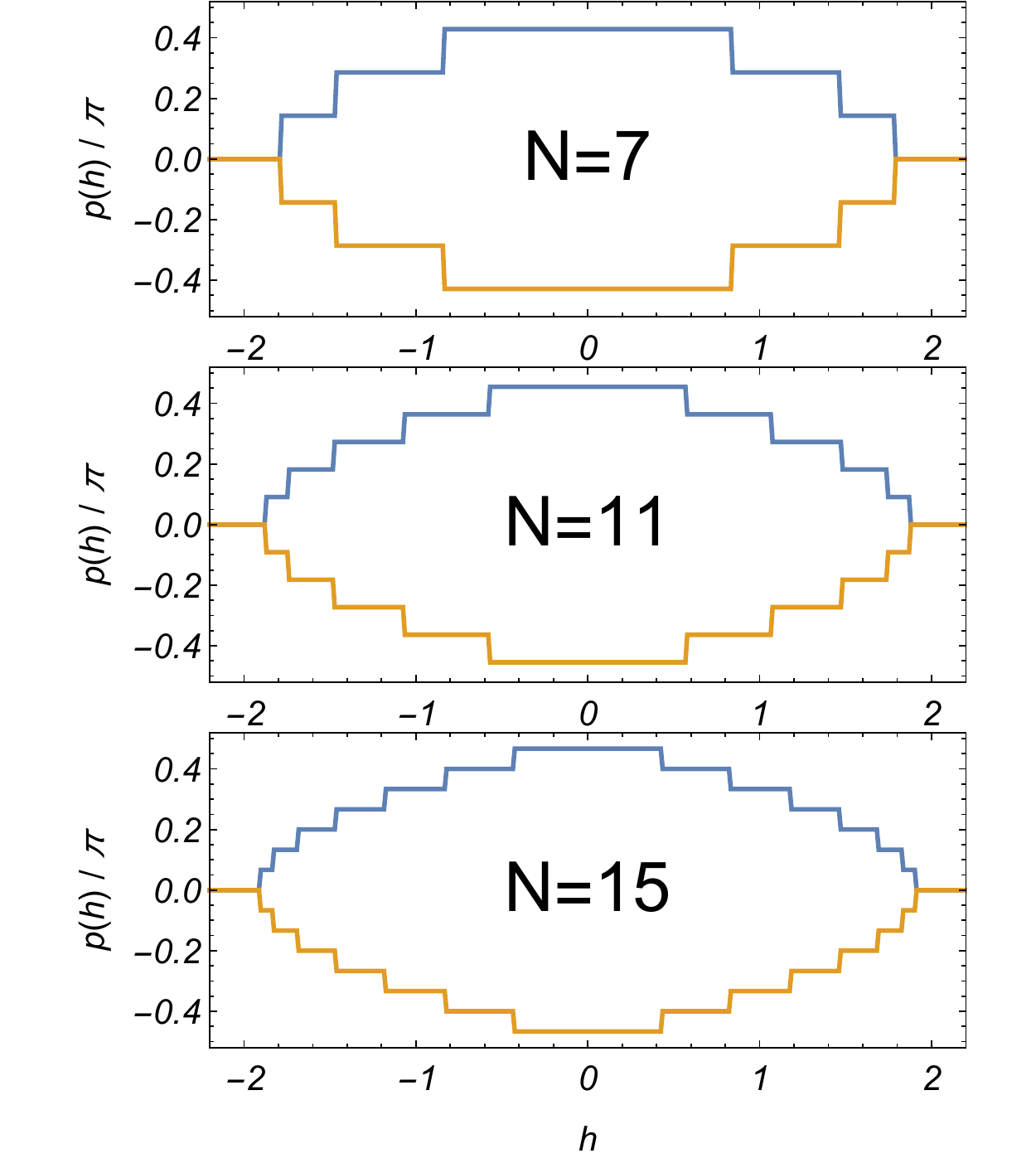}
\caption{Ground-state momenta $p(h)$ in unit of $\pi$ as a function of the local field amplitude $h$ for different sizes of the chain at fixed values of the anisotropies (note that momenta are compactified between $-\pi/2$ and $\pi/2$).
The data are obtained setting $J_x=1.0$, $J_y=0.6$ and $J_z=0.2$. 
Note that, except in the cases of a zero ground state momentum, the ground state manifold is always twofold degenerate with the two ground states carrying opposite momenta. Moving in phase space, the ground state vectors acquire all momenta allowed by the quantization rule.}
\label{fig:gsmomenta}
\end{figure}

{Despite its apparent simplicity, the model in \eqref{eq:Hamiltonian_ns} does not admit any analytical solution.}
Therefore, all our analyses are based on a numerical approach based on the Lanczos algorithm~\cite{Lanczos, Ojalvo} {for the exact diagionalization of the Hamiltonian}.   
Keeping $J_x=1$ and assuming $J_x>|J_y|,|J_z|$, we obtain that, for several odd $N$,  in the region $J_z>-J_y$ there is a threshold value of the local field $h^*>0$, which delimits the chiral region, such that for each $h\in(-h^*,h^*)$ the ground-state manifold is at least two-fold degenerate (see Fig.~\ref{fig:criticalfield}).
Such a manifold can be completely described in terms of the {eigenstates of the momentum operator $P$  that is the generator of the translation operator $T$, i.e. $T=e^{\imath P}$, whose action shifts all the spins by one site in the lattice. 
For a one-dimensional system with periodic boundary condition, the operator $T$ can be written in terms of the Pauli ones as~\cite{Vourdas2004,Franchini17}}
\begin{equation}
    T=\bigotimes_{i=1}^{N-1}\frac{1}{2}\bigg(\mathbb{I} + \sum_{\alpha=x,y,z}\sigma_i^\alpha\sigma_{i+1}^\alpha\bigg), 
\end{equation}
From the numerical analysis, we have that the ground state manifold admits a basis made by the two eigenstates of the momentum $\ket{\pm p}$ with opposite quantum number $\pm p(h)$, as it can be seen in Fig.~\ref{fig:gsmomenta}. 
Moreover, the momentum $p(h)$ acts as a Fermi point, in a system whose discrete symmetry should not allow for its existence, as low energy excitations lie nearby.
Moving $h$ throughout the region $(-h^*,h^*)$, the system visits all possible values of the momentum quantum number.
Since states with different quantum numbers are exactly orthogonal to each other and the size of the set of the momentum quantum numbers scales with the chain length, the number of different crossovers occurring in the region $(-h^*,h^*)$ increases with $N$.

\begin{figure}[t]
\includegraphics[width=0.95\columnwidth]{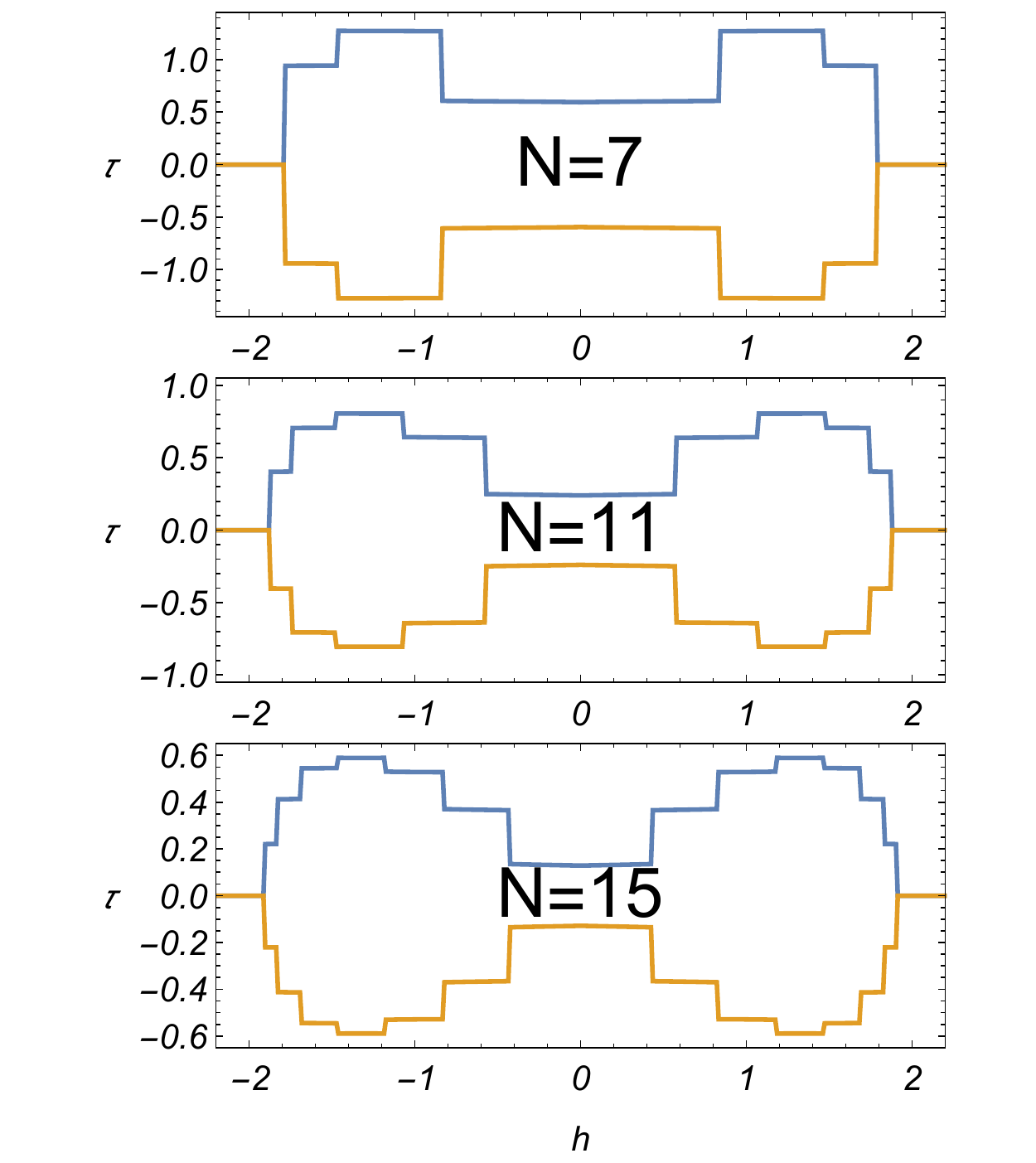}
\caption{Ground-state chirality $\tau$, evaluated on the elements of the ground-state manifolds that are also eigenstates of the lattice momentum operator. The results are plotted as a function of the local field amplitude $h$ for different sizes of the chain at fixed values of the anisotropies. A finite chirality reflects the finite momentum carried by the ground state vector, indicating that, in its lowest energy state, the system moves in a stationary way. 
The data are obtained settings $J_x=1.0$, $J_y=0.6$ and $J_z=0.2$.}
\label{fig:chiralorder}
\end{figure}

\begin{figure}[t]
\includegraphics[width=\columnwidth]{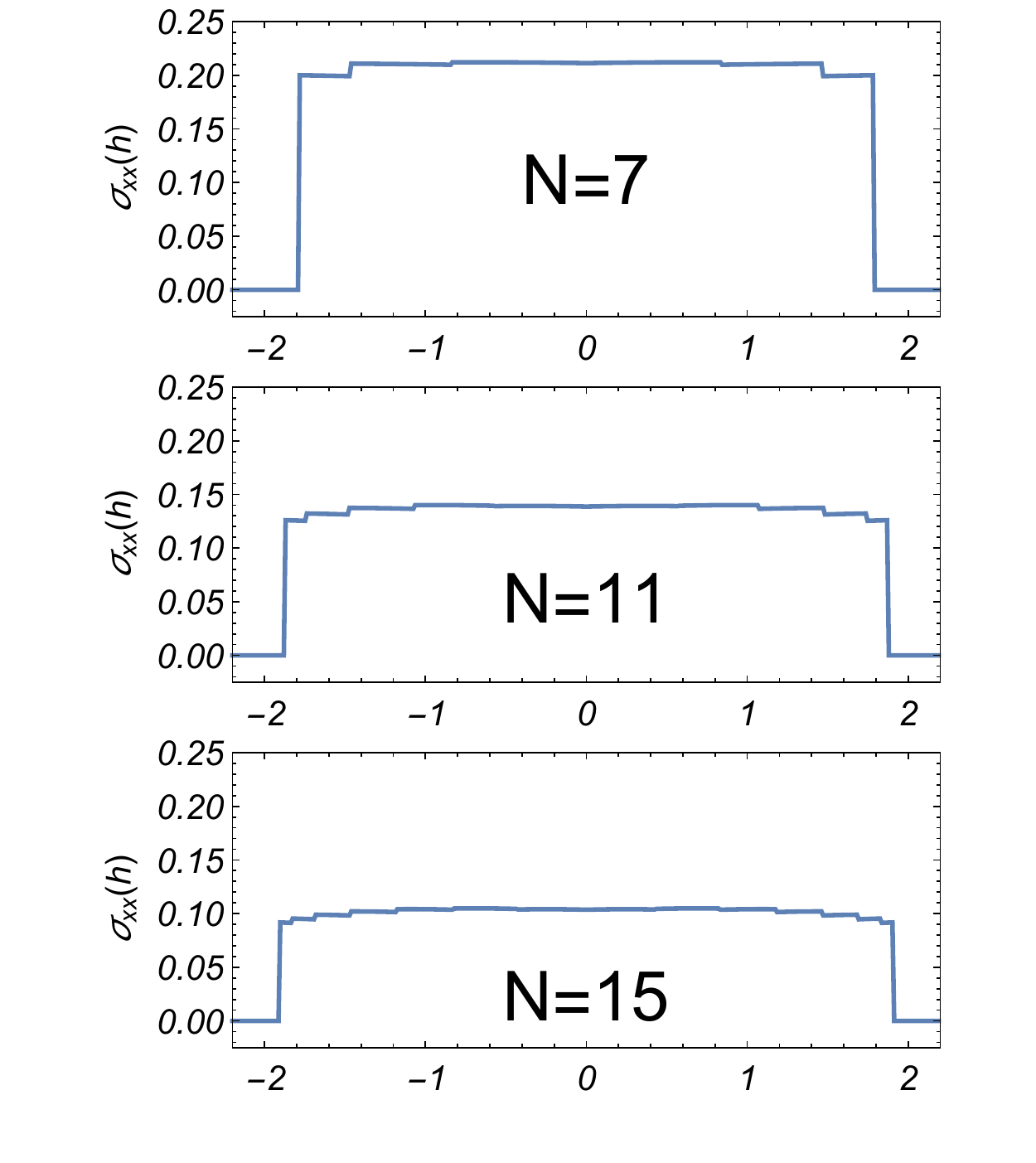}
\caption{Variance $\sigma_{xx}(h)$ of expectation values of $\sigma_i^x \sigma_{i+1}^x$, which varies in space, as function of the external field $h$ for $J_x=1.0$, $J_y=0.6$ and $J_z=0.2$. The data are obtained using the ground state vector $\frac{1}{\sqrt{2}}(\ket{p}+\ket{-p}$. The variance is finite for $|h|<h^*$, although decreasing with the chain length, indicating that indeed the two point function is not translational invariant.}
\label{fig:invariancebreaking}
\end{figure}

The presence of degeneracy in the ground state manifold of systems with only local discrete symmetries is not rare and it is at the basis of the mechanism of the symmetry violation. 
However, in the present case, we have two relevant peculiarities. 
First, the exact degeneracy is present even for finite sizes, while it usually appears, only in the thermodynamic limit. Second, the degeneracy involves states of the same parity, hence not allowing the system to violate the parity symmetry, as it happens in the ordered phase of unfrustrated systems.
On the contrary, such degeneracy stems from the mirror symmetry (that is, a reflection with respect to any site), which implies that each eigenstate has to be at least two-fold degenerate, with the only exception of zero momentum states~\cite{Maric2020_induced, Barwinkel}.

This structure of the ground-state manifold has some interesting implications.
Since in the chiral region $(-h^*,h^*)$ the ground-states typically have a non-vanishing momentum, they are expected to be characterized by a non-zero expectation value of the chiral operator $\hat{\tau}\equiv \overrightarrow{\sigma}_{i-1}\!\cdot\! \overrightarrow{\sigma}_{i}\! \times\! \overrightarrow{\sigma}_{i+1}$~\cite{Wen1989, Subrahmanyam1994}.
In Fig.~\ref{fig:chiralorder} we plot the behavior of the site independent expectation value $\tau \equiv \langle \pm p | \hat{\tau} | \! \pm p \rangle$. 
On the other side, a ground state that is a linear superposition of $\ket{\pm p}$, violates the invariance under the spatial translation of the Hamiltonian.
This fact {is not expected, at the ground-state level, in models possessing only discrete symmetries, and} can be highlighted by analyzing the variance of the spatial distribution of the two-body next-neighbor spin correlation function along $x$, i.e. the expectation values of $\sigma_i^x \sigma_{i+1}^x$, on the state $\frac{1}{ \sqrt{2}} \big( \ket{p}+\ket{-p} \big)$. 
Fixing $J_x=1$, $J_y=0.6$ and $J_z=0.2$ the variances as a function of the local field obtained with the exact diagonalization approach are depicted in Fig.~\ref{fig:invariancebreaking}.

{All these behaviors have already been previously observed in systems with continuous local symmetries, such as the Isotropic Heisenberg model~\cite{Subrahmanyam1994, Subrahmanyam, Poilblanc}, but they turn out to be completely new for systems with only local discrete symmetries.}

\section{Ground state fidelity}

{The analysis of the different quantities we have shown so far provides indirect information on the behavior of the ground state. 
{To access more directly the crossovers between the ground states} of the topological frustrated $XYZ$ chain, in this section, we will analyze a quantity derived from the quantum information theory, namely the ground state fidelity, which will give us a clearer picture of how the system responds to small changes in its parameters.}

For a parameters-dependent Hamiltonian $H(\overrightarrow{\lambda})$, the ground-state fidelity is defined as the square modulus of the overlap between two ground-states associated with slightly different sets of parameters, 
i.e.
\begin{equation}
    \mathcal{F(\overrightarrow{\lambda})}=~\left|\langle G(\overrightarrow{\lambda})|G(\overrightarrow{\lambda}+d\overrightarrow{\lambda})\rangle\right| \,.
    \label{eq:fidelity}
\end{equation}

{The ground state fidelity has already been widely used to analyze in detail the properties of the ground state of different one-dimensional systems \cite{Zanardi2006, Zanardi2007_1, Venuti2007, Zanardi2007_2, You2009, Kolodubretz2013, Bhattacharya2014}.}
{Since, in the thermodynamic limit, two neighboring ground state are always orthogonal, as a consequence of what is commonly referred to as the Anderson's Orthogonality Catastrophe~~\cite{Anderson1967-1, Anderson1967-2}, the most interesting quantity to consider is the finite size rate of change, the so called {\it Fidelity Susceptibility}. In systems with only discrete local symmetries, this quantity is continuous and diverges only approaching a quantum phase transition. It has thus been argued that it is a good quantity to use to detect criticality.}
On the contrary, in systems with continuous symmetry, {certain change in the system parameters trigger crossovers between states with different quantum numbers and the ground state fidelity presents as many points of discontinuity as the number of sites in the chain. For the $XXZ$ chain discussed above, these discontinuities are induced by changes in the external magnetic field and it represent an extreme case of orthogonality catastrophe because two neighboring ground states are exactly orthogonal even for small system sizes because they carry different quantum numbers.
In our case, the ground state fidelity behaves exactly like this:}
{moving in almost every direction in the parameter space, the system changes the ground-state momentum, making the fidelity drop to zero. But there exists a particular direction along which the momentum stays constant and the fidelity shows a more regular behavior. }

{To study how these properties scale with the system size, we need to consider much longer chains than those treatable within an exact diagonalization approach. 
Even advanced numerical methods, such as DMRG are not easily applicable to models showing topological frustration, especially in the region of interest of the present work, due to the simultaneous presence of both a degeneracy between ground-states in the same parity sector and the closing of the energy gap between the ground-states and the immediately overlying excited states.
Therefore, to push our study towards larger sizes we now focus on the case $J_z = 0$ which can be mapped into a free-fermionic model, thus disclosing the possibility for an analytical solution.
Introducing the anisotropic parameter $\gamma$ and defining $J_x=\frac{1+\gamma}{2}$ and $J_y=\frac{1-\gamma}{2}$, the Hamiltonian of the model in \eqref{eq:Hamiltonian_ns} reduces to the one of the $XY$ chain in a transverse field 
\begin{equation}
\label{eq:Hamiltonian}
    H=\sum_{j=1}^N \bigg[\frac{1+\gamma}{2}\sigma^x_{j}\sigma^x_{j+1}+\frac{1-\gamma}{2}\sigma^y_{j}\sigma^y_{j+1}-h\sigma^z_{j}\bigg]\, .
\end{equation}
{This is a prototypical exactly solvable model which, through a series of exact, although non-local, transformations can be brought into a free fermionic form. Its method of solution is known since the famous 1961 paper by Lieb, Schutz, and Mattis~\cite{LSM-1961} and since then many interesting observables have been calculated for it, including the fundamental correlation functions~\cite{Barouch1971}, establishing it as a corner stone in many-body, strongly correlated quantum systems. However, virtually all these works have been interested in bulk properties deemed independent from the boundary conditions and have thus been quite cavalier in this respect (notice that, without a proper account of boundary conditions it is not possible even to establish the asymptotic ground state degeneracy of this model in its ordered phase~\cite{Franchini17,kitaev2000}). As a matter of fact, as it is has been only recently appreciated, FBC induce several subtle differences in the solution of the $XY$ chain which yield surprising outcomes~\cite{Dong2016,Giampaolo2018, Maric2020_destroy,Maric2020_induced}, such as those that we discuss in this work. The exact analytical solution of this model is presented in the Appendix, while here we focus on the physical results.} {In complete agreement with the previous numerical results}, we find that it exists a chiral region, defined for $|h|<h^*=1-\gamma^2$, were the ground-state possesses a non-zero momentum $\tilde{q}$ which depends on the driving parameters as
\begin{equation}
    \tilde{q}=\arccos\bigg({\frac{h}{\gamma^2-1}}\bigg).
\end{equation}}

In this situation, we have two different possibilities for ground-state fidelity: {the states $|G(\overrightarrow{\lambda})\rangle$ and  $|G(\overrightarrow{\lambda}+d\overrightarrow{\lambda})\rangle$ can either be exactly orthogonal, or not. The first happens either} because they live in different parity sectors and hence are characterized by a different number of fermions, or because the fermions occupy different fermionic modes.

{Instead, if we move the parameter along one of the parabolas
$ h = \mathrm{c} (1 - \gamma^2) $, the two ground states have the same parity and the and their fidelity can be written as} (see Appendix for details)
\begin{equation}
\label{eq:fidelity_even}
    \mathcal{F}=\!\!\!\!\!\!\!\! \prod_{q\in\Gamma^+_2/\{\tilde{q}^+\}} \!\!\!\!\!\!\!\! \cos(\Tilde{\theta}_q-\theta_q) ,
\end{equation}
for the even parity and
\begin{equation}
\label{eq:fidelity_odd}
    \mathcal{F}= \!\!\!\!\!\!\!\! \prod_{q\in\Gamma^-_2/\{\tilde{q}^-\}}\!\!\!\!\!\!\!\!\cos(\Tilde{\theta}_q-\theta_q)\,,
\end{equation}
for the odd one. These expressions are 
very similar 
to the ones characterizing the ground-state fidelity of both the unfrustrated systems and the region with $|h|>h^*$.

\begin{figure}
\centerline{\includegraphics[width=0.9\columnwidth]{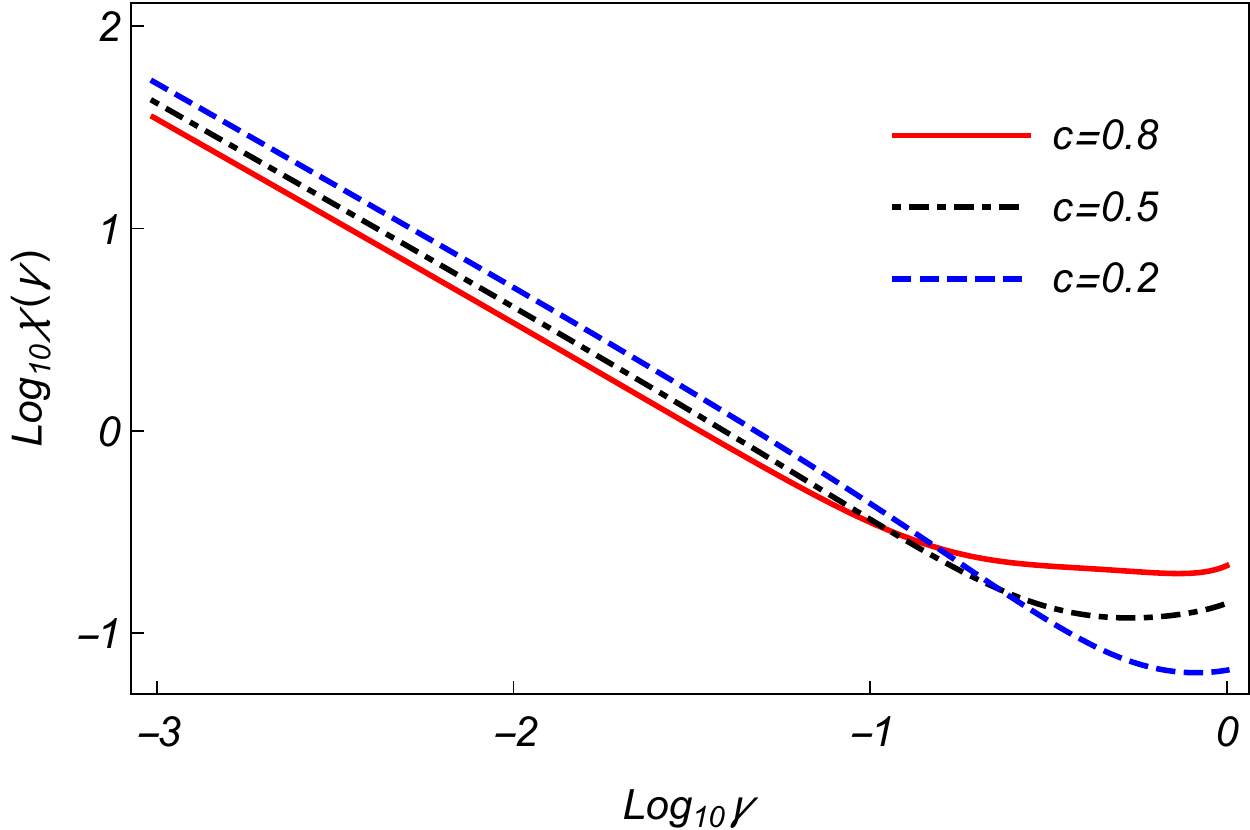}}
\caption{Thermodynamic limit of the global state fidelity susceptibility $\tilde{\chi}$ along manifolds of constant momentum $\tilde{q}$ obtained moving along the parabola $h=\mathrm{c}(1-\gamma^2)$ as a function of $\gamma$. Going towards $\gamma=0$ the fidelity susceptibility diverges signaling the presence of the quantum phase transition between the anisotropic $XY$ chain and the isotropic $XX$ one.}
\label{fig:global_susc}
\end{figure}

Along these parabolas it is also possible to evaluate the fidelity susceptibility $\chi$ that, by definition, is equal to the leading order of the expansion of the ground-state fidelity in the parameter change:
\begin{equation}
    \mathcal{F}\approx 1-\frac{1}{2} \chi \, d\gamma^2,
    \label{eq:fidelsusc}
\end{equation}
Such quantity has been widely studied in the context of the unfrustrated \textit{XY} chain~\cite{Zanardi2007_1, Venuti2007}, proving to be able to correctly predict the phase transition at $h=\pm 1$ and $\gamma=0$~\cite{Kolodubretz2013}. 
In agreement with the Anderson orthogonality catastrophe~\cite{Anderson1967-1, Anderson1967-2}, the fidelity susceptibility always tends to diverge in the thermodynamic limit. 
However, while at a regular point, this divergence is only extensive, it becomes super-extensive close to a quantum phase transition~\cite{Venuti2007}. 
Therefore, to study the behavior of the fidelity susceptibility in the thermodynamic limit let us introduce the re-normalized fidelity susceptibility $\tilde{\chi}$ obtained by dividing $\chi$ by the system volume. 
After a long but straightforward evaluation it is possible to prove (see Appendix for details) that, independently from the parity sector, the normalized fidelity susceptibility is
\begin{equation}
    \tilde{\chi}=\frac{1}{16}\frac{1+c^2(1+\gamma)^3(3\gamma-1)}{\gamma(1+\gamma)^2(1-c^2(1-\gamma^2)^2)}.
    \label{tildechi}
\end{equation}
One can check, and Fig.~\ref{fig:global_susc} confirms, that the re-normalized fidelity susceptibility diverges at $\gamma\to0$, hence signaling the presence of the critical phase of the quantum $XX$ chain, i.e. the continuous symmetry model emerging by setting $\gamma=0$.

However, as noted above, if we do not assume that $ \overrightarrow{\lambda} $ and $ \overrightarrow{\lambda} + d \overrightarrow{\lambda} $ are on the same parabola, the ground state fidelity vanishes identically as it occurs within the gapless phases of systems with continuous symmetries.
In the latter case, to avoid the problem associated with the scaling analysis of such a ground-state fidelity, it is normal to resort to the reduced fidelity~\cite{Kwok2008, You2009, Gu2008, Ma2008}.
The reduced fidelity can be seen as a generalization of the ground state fidelity, and it represents the overlap between the reduced density matrices for a fixed subsystem, obtained from the ground states corresponding to different parameters, i.e.
\begin{equation}
    \mathcal{F}_{R_A}=\Tr \sqrt{\rho_A(\vec{\lambda})^{1/2}\rho_A(\vec{\lambda}+\vec{d\lambda})\rho_A(\vec{\lambda})^{1/2}} \; .
\end{equation}
Here $\rho_A(\vec{\lambda})$  ($\rho_A(\vec{\lambda}+\vec{d\lambda})$) denotes the reduced density matrix of the ground state $|G(\vec{\lambda})\rangle$ ($|G(\vec{\lambda}+d\vec{\lambda})\rangle$), obtained by tracing out all the degrees of freedom associated to sites outside the chosen subset $A$: {$\rho_A(\vec{\lambda}) \equiv \tr_B |G(\vec{\lambda})\rangle \langle G(\vec{\lambda})| (\rho_A(\vec{\lambda}+\vec{d\lambda}) \equiv \tr_B |G(\vec{\lambda}+\vec{d\lambda})\rangle \langle G(\vec{\lambda}+\vec{d\lambda})|) $.}
Among all the possibilities we decided to focus on the reduced matrix obtained by projecting the ground state on two nearest-neighbor spins, but {we checked that} other choices lead to similar results.
The reduced density matrix on two nearest-neighbor sites can be written in terms of the spin correlation functions~\cite{Osborne2002} as 
\begin{equation}
\label{eq:reduced_fidelity}
    \rho_{ij}=\frac{1}{4}\sum_{\alpha,\beta=0,x,y,z}\langle \sigma_i^\alpha \sigma_j^\beta \rangle \; \sigma_i^\alpha \otimes \sigma_j^\beta,
\end{equation}
where $\sigma^0$ denotes the identity and the analytic expressions for the correlation functions appearing in \eqref{eq:reduced_fidelity} are presented in the Appendix.

\begin{figure}[t]
\centerline{\includegraphics[width=0.9\columnwidth]{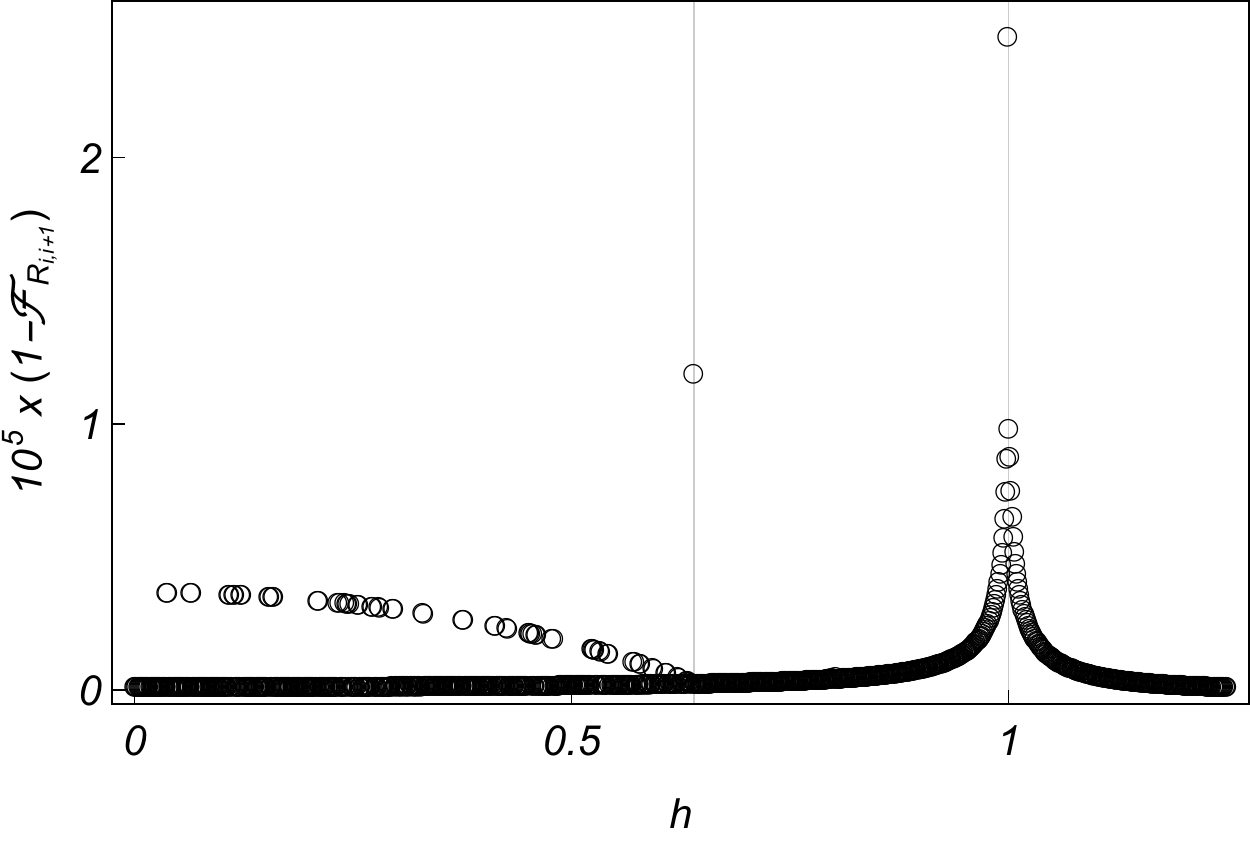}}
\caption{Reduced fidelity obtained projecting a ground state eigenstate of the momentum operators into the Hilbert space defined by two nearest neighbor spins. The data are obtained considering and \textit{XY} spin chain made of $1001$ spins, fixing $\gamma=0.6$ and moving $h$ from 0 to $1.25$. For $h<h^*$ the lower points refer to movement along a parabola $h=\mathrm{c}(1-\gamma^2)$, while the higher values represent a generic flow in which neighboring ground states are characterized by different occupied modes.}
\label{fig:red_fid_1}
\end{figure}

The results obtained for the reduced fidelity, in a system composed of $1001$ sites, by keeping the value of $\gamma $ fixed and changing $ h $ with uniform steps equal to $ 10^{- 4} $ are shown in Figure~\ref{fig:red_fid_1}.

\begin{figure}[t]
\centerline{\includegraphics[width=0.9\columnwidth]{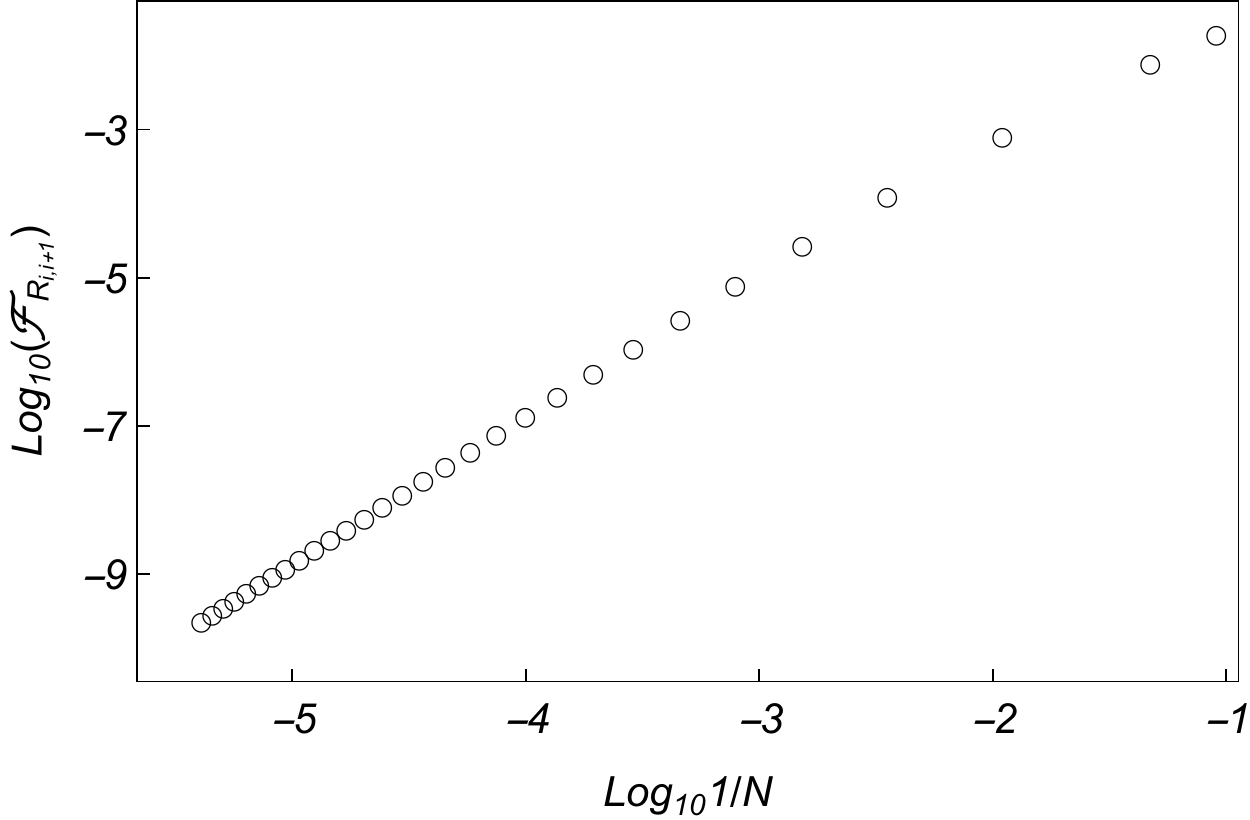}}
\caption{Reduced fidelity at the line $h=h^*$ obtained projecting a ground state eigenstate of the momentum operators into the Hilbert space defined by two next neighbor spins as function of the size of the system. 
The data are obtained considering $\gamma=0.6$. Although at finite sizes a discontinuity is evident, in the thermodynamic limit it vanishes algebraically.}
\label{fig:red_fid_2}
\end{figure}

In the chiral region $h<h^*$ we plot two sets of points: the lowest ones refer to the reduced fidelity while moving along one of the parabolas which keep the occupied modes in the ground state fixed, while the higher ones represent a generic change for which neighboring ground states have vanishing overlaps and even the reduced fidelity gets significantly dampened. Note that a clear discontinuity is observable at the boundary of the chiral region for $h=h^*$, where an isolated point develops, which reflects the fact that outside the chiral region the ground state is a vacuum and thus has different correlation functions compared to those for $h<h^*$.
However, in Fig.~\ref{fig:red_fid_2} we can observe that the value of this discontinuity decreases proportionally to $ 1 / N^2 $ and therefore disappears in the thermodynamic limit.
Similar analysis can be performed for all other points in which the reduced fidelity shows a discontinuity, always yielding discontinuities which vanish algebraically with the chain length and hence, in the thermodynamic limit, the behavior of the reduced fidelity for the topologically frustrated spin models is indistinguishable from the one of the unfrustrated models. We should remark, here, that in systems with continuous symmetry, {although the discontinuities in the reduced fidelity susceptibility between neighboring state also vanish in the thermodynamic limit,} the region of crossovers between different ground states is a true quantum phase and thus the discontinuity at the boundary survives the thermodynamic limit~\cite{Kwok2008}.

Clearly, since on one side the whole ground state fidelity is singular and produces a foliation of the phase space, while the two sites reduced fidelity becomes continuous in the thermodynamic limit, a crossover is expected between two behaviors if more sites are included in the subset $A$, ideally scaling with the total chain length. However, such analysis cannot be carried out analytically and requires too heavy of a numerical study, which is beyond the scope of the current work.

\section{Characterization of the critical line}

{Thanks to the analytical solution of the $XY$ chain it is possible to study how all the other features of the chiral region scale in the thermodynamic limit, namely the chirality and the breaking of translational invariance. Hence, we can address the question of whether this represents a different thermodynamic phase or not.}

\begin{figure}[t]
\includegraphics[width=0.9\columnwidth]{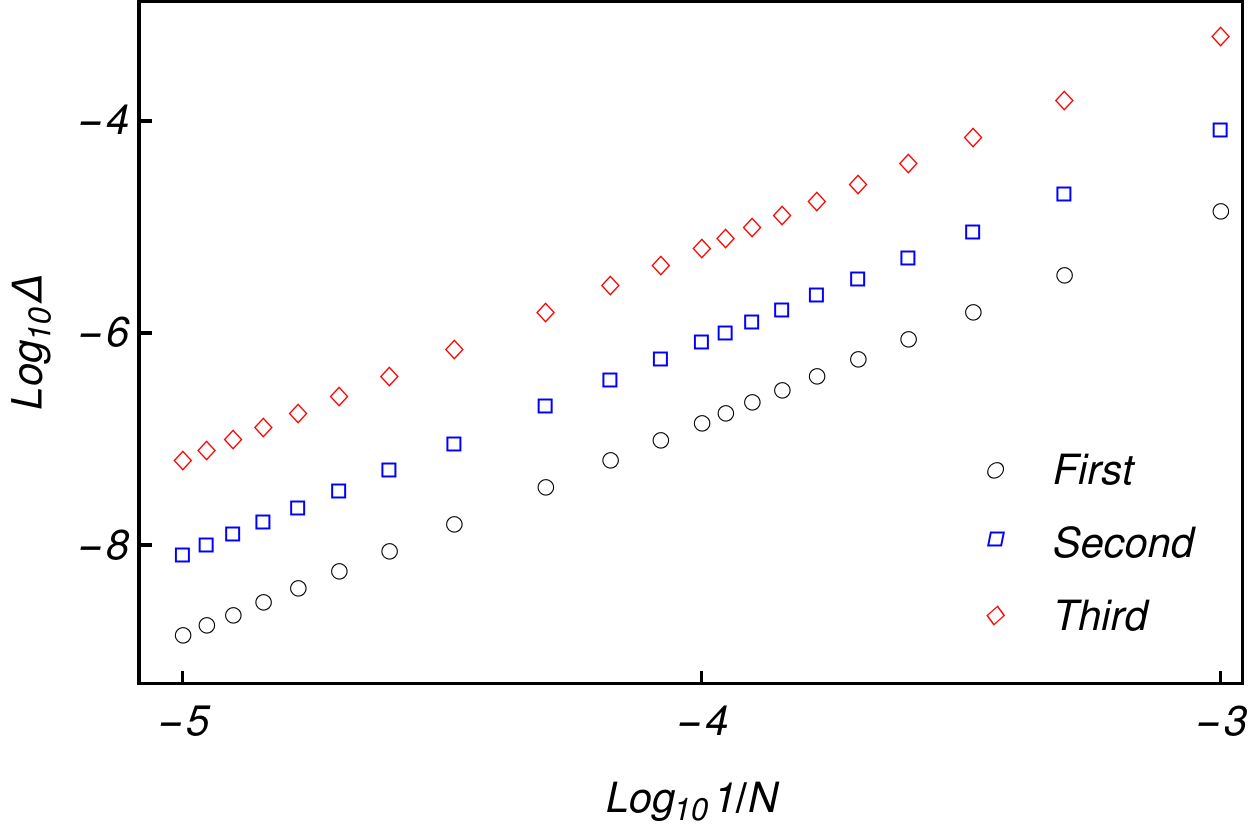}
\caption{Behavior of the jump in the energy derivatives $\Delta(n)$, for $n=1,\,2,\,3$ as a function of the inverse chain length, in logarithmic scale, obtained crossing the threshold $h=h^*$ along a line with constant $\gamma$. The data are obtained setting  $\gamma=0.6$. All derivatives display a vanishing discontinuity in the thermodynamic limit.}
\label{fig:diff_derivative}
\end{figure}

For $h^* <h <1$ one can show that the ground state is always represented by the fermionic vacuum state, while in the chiral region the ground state manifold keeps changing its parity and momenta. Even increasing the chain length without moving $h$ and $\gamma$ can switch the ground state parity, {because of the shift in the momenta quantization}. Moreover, the gap between the alternating ground-states in different parity sectors closes exponentially with the chain length, which means that in the thermodynamic limit the two manifolds become effectively degenerate: crossing the line $h=h^*$ the ground state degeneracy thus grows from $1$ to $4$, which could indicate a first-order quantum phase transition.
However, analyzing the free energy derivatives (which at zero temperature coincide with the ground state energy) we cannot detect any discontinuity that remains finite in the thermodynamic limit, as shown in Fig.~\ref{fig:diff_derivative}.
This fact implies that in the thermodynamic limit all derivatives are analytical and hence that if $h=h^*$ represents a quantum phase transition, it has to be one akin to a BKT transition~\cite{BKT1, BKT2, BKT3}.
This result is in stark contrast 
with the other phase transition induced by topological frustration discovered in models without external field~\cite{Maric2020_induced}. 
In fact, in that transition, the first derivative of the ground-state energy shows a discontinuity that stays finite even in the thermodynamic limit.
The reason behind this different behavior is that the transition point in  Ref.~\cite{Maric2020_induced} possesses a higher symmetry that produces a massive (thermodynamically large) ground state degeneracy. In this way, crossing this point, there is a true discontinuity and, for instance, the mode occupied in the odd parity sector of the ground state has momentum close to $\pm \frac{\pi}{2}$, instead of $\pm \pi$ as we have here.

On the other hand, phase transitions are associated with a macroscopic reordering of the system properties that can be detected by opportunely chosen quantities.
From what we have seen in the previous section, among others, two possible quantities can be considered: the chirality parameter that detects the existence of ground-states with a non-vanishing momentum and the spatial variance of local observables that highlights the violation of the invariance under spatial translation. 
Both quantities, and in general each spin correlation function on a ground-state with a fixed parity, can be obtained analytically in the framework of the analytical approach that we are using.  
The key point is the introduction of two sets of Majorana operators so that each spin correlation function can be mapped to a string of Majoranas~\cite{Franchini17}.
By exploiting Wick's theorem, the expectation value of these correlation functions can be reduced to the evaluation of Pfaffians whose elements are the expectation values of two Majorana operators. 
As we show in the appendix, for $h>h^*$ these expectation values can be classified into two different families: a) when the two Majorana operators come from the same set, the expectation value vanishes unless the two operators coincide; b) when the two Majorana operators come from different sets, the expectation value assumes values in the range $[-1,1]$ and are invariant under spatial translation.
Entering the region $h<h^*$ both these properties are changed.
Indeed, the expectation values of Majorana operators coming from the same set but defined on different fermionic sites assume values proportional to $1/N$ that hold the property to be invariant under spatial translation.
On the other hand, the expectation values for Majorana in different sets acquire corrections proportional to $1/N$ that explicitly violate the invariance under spatial translation.

These two corrections and their proportionality to $1/N$ explain why both the chirality and the variance of the distribution of local quantities are different from zero in a finite frustrated system but disappear when the thermodynamic limit is taken into account.
Indeed, using Wick's theorem, the expectation value of the chirality operator $\tau$ can be reduced to a sum of products of expectations of pairs of Majorana fermions, with the peculiarity that each term in the sum contains at least an expectation on pairs of Majorana belonging to the same set, thus providing an algebraic decay with the system size to the whole expression.
On the other hand, since all the site-dependent contributions to a local quantity scale with $1/N$, the variance is also vanishing in the thermodynamic limit. 
Both these behaviors should be compared to other observables calculated in presence of topological frustration in ~\cite{Maric2020_destroy,Maric2020_induced}: there the $1/N$ corrections coming from frustration appeared in combination with finite terms present also in absence of frustration and when an expectation value involved a sufficiently high number of corrections (also scaling with $N$), the resulting expression brought finite contributions.
\begin{figure}[t]
\includegraphics[width=0.9\columnwidth]{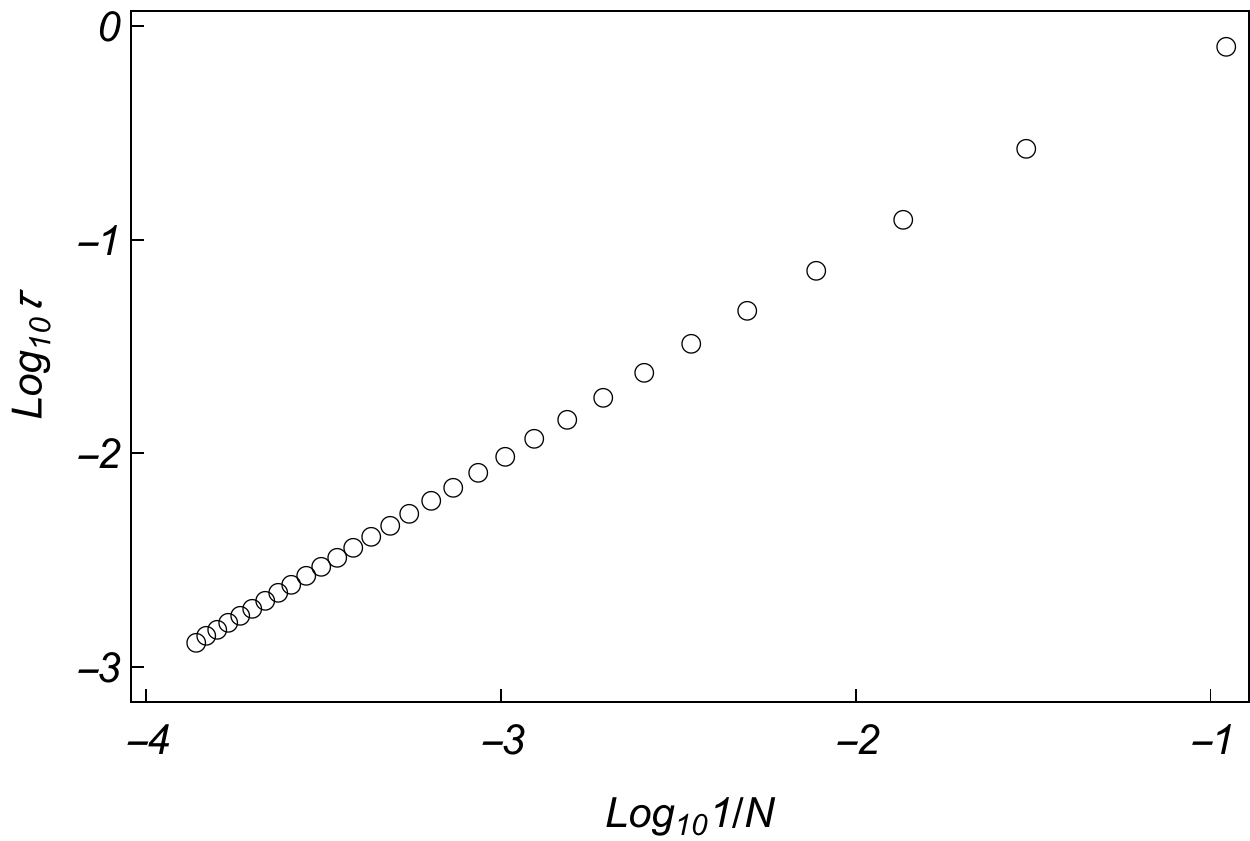}
\caption{Behavior of the chirality $\tau$ for ground states that are eigenvalues of the momentum operator as a function of the length of the chain $N$. The data are obtained for the \textit{XY} chain setting $\gamma=0.6$ and $h=0.4$ }
\label{fig:xy_chirality}
\end{figure}
\begin{figure}[b]
\includegraphics[width=0.9\columnwidth]{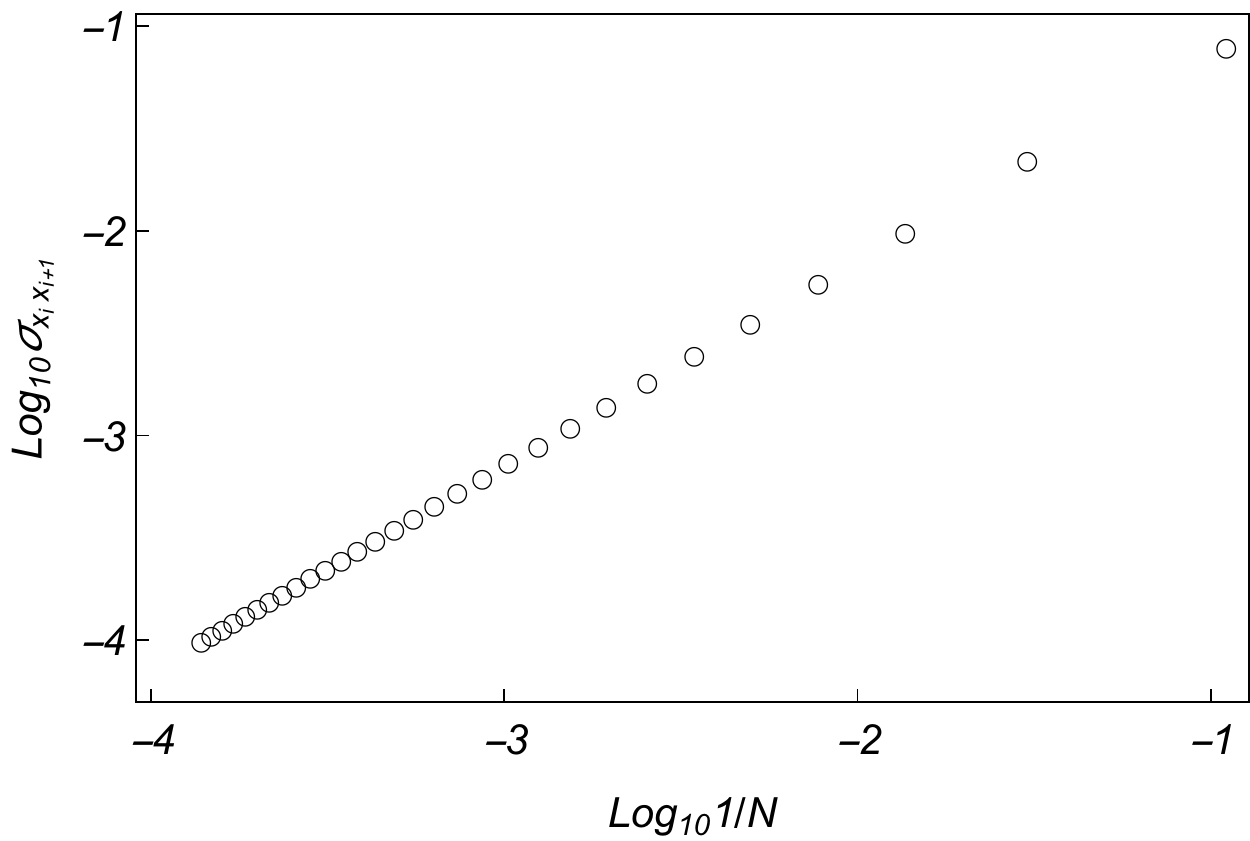}
\caption{Variance of the spatial distribution for the two-point spin correlation functions $\langle\sigma^x_i\sigma^x_{i+1}\rangle$ for ground states obtained as a real symmetric combination of the two ground states with a definite momentum, as a function of the length of the chain $N$. The data are obtained for the \textit{XY} chain setting $\gamma=0.6$ and $h=0.4$}
\label{fig:xy_invariance_breaking}
\end{figure}
The analytic derivation of both quantities can be found in the Appendix while in Fig.~\ref{fig:xy_chirality} and Fig.~\ref{fig:xy_invariance_breaking} their dependence on the chain length is depicted for some relevant ground-states choice.
The disappearance in the thermodynamic limit of both properties that characterize this region of the parameter space, associated with the absence of any local discontinuity in the energy derivatives on the line $h=h^*$ support the idea that we are looking at a boundary transition whose effects disappear when the size of the system diverges.

\section{Conclusions}

In conclusion, we have seen how, in presence of topological frustration, an anisotropic Heisenberg chain, which presents only discrete rotational symmetries associated with finite sets of quantum numbers, is characterized by a region of parameter space ({\it chiral region})in which the system mimics that of a system with continuous symmetries. 
In fact, in analogy with the latter, the system not only presents a gapless energy spectrum with a finite Fermi momentum but shows, in the thermodynamic limit, a ground-state characterized by a continuous cross-over between two-dimensional mutually orthogonal manifolds. 
Each one of these manifolds is spanned by eigenstates of the lattice momentum with the same eigenvalues but opposite signs. 
This fact has several interesting consequences.
At first, since these states are characterized by a non-zero momentum when the system is made of a finite number of spins, they show a non-zero chirality that vanishes in the thermodynamic limit. 
At the same time, any non-trivial linear combination of such states produces a new ground-state violating the invariance under spatial translation, which, in the case of finite systems, can be observed through the variance associated with the spatial distribution of local observables, but which are zeroed when the dimension of the system diverges.
This chiral region is separated from the rest by a threshold line that separates it from a region in which the system is still gap-less but the ground state is unique and characterized by zero momentum. 
Our analysis clearly shows that such change in the ground state degeneracy is not mirrored in the behavior of the energy. 
Indeed, in the thermodynamic limit, all energy derivatives are analytical, and hence the transition has to be akin to a BKT transition. 
However, the disappearance in the thermodynamic limit of both properties that characterize this region of parameter space supports the idea that we are looking at a boundary transition whose effects disappear when the size of the system diverges. {Indeed, to the best of our knowledge, boundary BKT transitions have not been observed before.}

Consistently with this picture, the ground state fidelity obtained by continuously varying the parameters of the system, in the thermodynamic limit, is identically zero in almost all directions. The only exception is obtained when the change of the Hamiltonian parameters is carried out in such a way as to keep the momenta characterizing the ground state manifold constant, which, in the particular case of the $XY$ chain for which it is possible to carry out an analytical treatment, occurs when the ratio $ h / (1- \gamma^2) $ is kept constant.

However, despite these results, several points concerning this new region of the parameter space of topologically frustrated systems still remain unanswered. For example, it is not yet clear whether it is possible to characterize this new phase from the topological point of view. Most of all, it is important to test the resilience of the described phenomenology to the presence of localized defects, in view of its possible exploitation for quantum technologies. These and other topics will be dealt with in future papers which are already in preparation.

\section*{Acknowledgments}

We thank  Vanja Mari\'c for numerous discussions and for the preliminary work from which the idea for this work stemmed.
A. G. Catalano acknowledges support from the MOQS ITN programme, a European Union’s Horizon 2020 research and innovation program under the Marie Skłodowska-Curie grant agreement number 955479..
S.M.G and F.F. acknowledge support from the Croatian Science Funds Project No. IP-2019–4–3321 and
the QuantiXLie Center of Excellence, a project co-financed by the Croatian Government and European
12 Union through the European Regional Development Fund–the Competitiveness and Cohesion Operational Programme (Grant KK.01.1.1.01.0004).

\appendix

\section{Solution of the topologically frustrated $XY$ chain}

The \textit{XY} chain in equation~\eqref{eq:Hamiltonian} can be diagonalized exactly.
For the sake of simplicity, we limit our analysis to the case with $h\ge 0$ and $0<\gamma\le1$, but our results can be easily extended also to the other regions of parameters space. 
The standard procedure prescribes a mapping of spin operators into fermionic ones, which are defined by the Jordan-Wigner transformation:
\begin{equation}
	\label{eq:JW-a}
	\sigma^-_j=\prod_{l<j} \sigma^z_l\psi_l^\dagger, \quad \sigma^+_j=\prod_{l<j} \sigma^z_l\psi_j, \quad \sigma^z_j=1-2\psi_j^\dagger\psi_j,
\end{equation}
where $\psi_l$ ($\psi^\dagger_l$) are fermionic annihilation (creation) operators.
In terms of such operators, taking into account the periodic boundary conditions and discarding constant terms, the Hamiltonian thus becomes
\begin{eqnarray}
	H\!&\!=\!&\!\sum_{j=1}^{N-1}\!\bigg[\psi^\dagger_{j+1}\psi_j\!+\!\psi^\dagger_{j}\psi_{j+1}\!+\!\gamma(\psi^\dagger_{j}\psi^\dagger_{j+1}\!+\!\psi_{j+1}\psi_j) \bigg]  \\
	\! &\! +\! &\! 2h\sum_{j=1}^N\psi^\dagger_j\psi_j \!+\! \Pi^z \bigg[\psi^\dagger_{1}\psi_N\!+\!\psi^\dagger_{N}\psi_{1}\!+\!\gamma(\psi^\dagger_{N}\psi^\dagger_1\!+\!\psi_{1}\psi_N)\bigg].\nonumber
\end{eqnarray}
The latter expression is not quadratic itself, but reduces to a quadratic form in each of the parity sectors of $\Pi^z$. Therefore, it is convenient to write it in the form  
\begin{equation*}
	H=\frac{1+\Pi^z}{2}H^+\frac{1+\Pi^z}{2}+\frac{1-\Pi^z}{2}H^-\frac{1-\Pi^z}{2},
\end{equation*}
where both $H^\pm$ are quadratic. 
Hence we can bring the Hamiltonian into a free fermion form by means of two final steps. 
First, we perform a Fourier transform
\begin{equation}
	\psi_q=\frac{e^{-\imath \pi/4}}{\sqrt{N}}\sum_{j=1}^N e^{-\imath qj}\psi_j.
\end{equation}
It is worth noting that, due to the different quantization conditions, $H^\pm$ are defined on two different sets of fermionic modes, respectively $q\in\Gamma^-=\{\frac{2\pi n}{N}\}_{n=0}^{N-1}$ in the odd sector and $q\in\Gamma^+=\{\frac{2\pi}{N}(n+\frac{1}{2})\}_{n=0}^{N-1}$ in the even one.
Finally a Bogoliubov rotation in Fourier space 
\begin{equation}
	\label{eq:hi-a}
	b_q=\cos \theta_q \psi_q + \sin \theta_q \psi^\dagger_{-q}, 
\end{equation}
with momentum-dependent Bogoliubov angles 
\begin{equation}
	\label{eq:bogoliubov_angles-a}
	\theta_q=\frac{1}{2}\arctan\bigg(\frac{\gamma \sin q}{h+\cos q}\bigg) \;\; q\neq0,\pi\;\;,\;\; \theta_{0,\pi}=0,
\end{equation}
leads to the Hamiltonians
\begin{subequations}
\begin{eqnarray}
    \label{eq:ham-}
    H^-\!&\!=\!&\!\!\!\!\!\!\!\sum_{q\in \Gamma^-/\{0\}}\!\!\!\!\!\!\Lambda(q)\bigg(b^\dagger_q b_q\!-\!\frac{1}{2}\bigg) +\epsilon(0)\bigg(b^\dagger_0 b_0\!-\!\frac{1}{2}\bigg)\\
    \label{eq:ham+}
    H^+\!&\!=\!&\!\!\!\!\!\!\!\sum_{q\in \Gamma^+/\{\pi\}}\!\!\!\!\!\!\Lambda(q)\bigg(b^\dagger_q b_q\!-\!\frac{1}{2}\bigg) +\epsilon(\pi)\bigg(b^\dagger_\pi b_\pi\!-\!\frac{1}{2}\bigg),
\end{eqnarray}
\end{subequations}
Here $b_q$ ($b^\dagger_q$) are the Bogoliubov annihilation (creation) fermionic operators.
The dispersion relation $\Lambda(q)$ for $q\neq0,\,\pi$ obeys
\begin{equation}
\label{eq:lambda}
    \Lambda(q)=\sqrt{(h+\cos q)^2+\gamma^2\sin^2q}, 
\end{equation}
while for the two specific modes $q=0\in\Gamma^-$ and  $q=\pi\in\Gamma^+$ we have 
\begin{equation}
    \epsilon(0)=h+1, \quad \epsilon(\pi)=h-1.
\end{equation}
{Having brought the Hamiltonian in the form of a free fermionic system, although with a non-trivial single particle spectrum, we can construct the Hilbert space of the spin chain in terms of the Fock space of the fermionic one, starting from the vacuum states $\ket{\emptyset^\pm}$, defined separately in each parity sector as the states annihilated by all the corresponding destruction operators: $b_q \ket{\emptyset^\pm} = 0, \forall q \in \Gamma^\pm$, and by applying creation operators, always respecting the parity constraint.}

\begin{figure}[t]
	\centerline{\includegraphics[width=0.9\columnwidth]{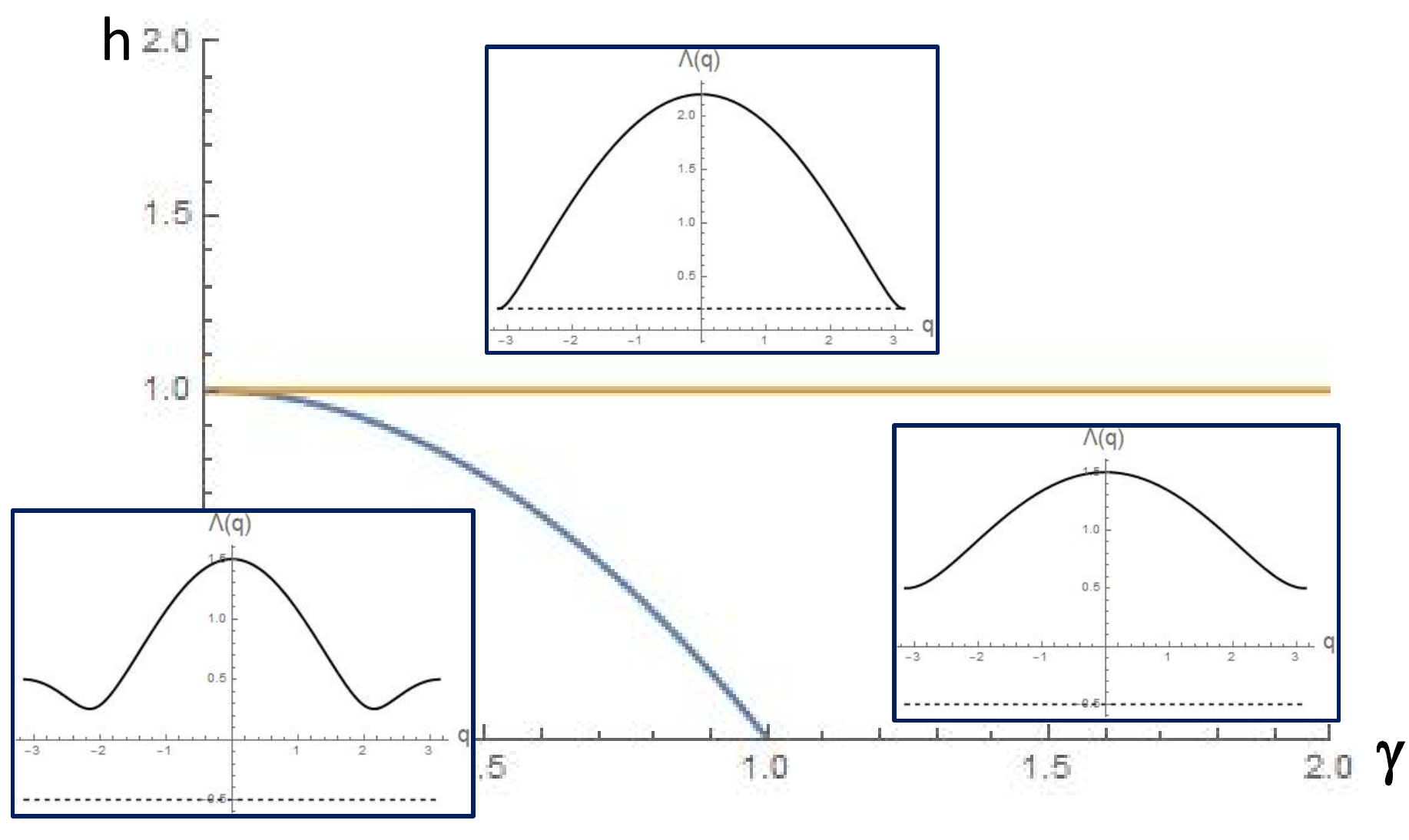}}
	\caption{Phase diagram in the $(\gamma,h$) space of the frustrated $XY$ chain: the line at $h=1$ is the phase transition separating the frustrated phase from a phase where FBC do not affect the behavior of the system. The parabola $h=1-\gamma^2$ separates the chiral region from the region with a unique ground state. Instances of the single particle dispersion relation eq.~\eqref{eq:lambda} are plotted in the various regions together with the energy of the $\pi$ mode (dashed line), whose occupation lowers the energy of the system for $h<1$.}
	\label{fig:phasediag}
\end{figure}

It is important to observe that, having assumed $h>0$, all fermionic modes in the odd sector are associated with a positive energy, while in the even sector 
there is also a (single) negative contribution to the total energy coming from a fermion with momentum $q=\pi$ if $h<1$.
As depicted in Fig.~\ref{fig:phasediag}, we can then partition the phase diagram with $h \ge 0 $ and $0\le \gamma\le 1$ into three regions: 
\begin{itemize}
	
	\item[\textbf{I}]
	$h>1$:  
	In this case, every excited mode brings a positive energy to the system and hence the lowest energy state is the Bogoliubov vacuum state $\ket{\emptyset^+}$ falling in the even parity sector. 
	Such a state is separated from the rest of the spectrum by a finite amount of energy equal to the minimum energy carried by the presence of a single Bogoliubov fermion in the system.
	
	\item[\textbf{II}] $1-\gamma^2\le h\le1$: In this region, the presence of a fermion in the $\pi$-mode, differently from all other modes in the system, is associated with a negative contribution to the total energy. 
	Hence, when populated, the $\pi$-mode lowers the energy of the system by a finite amount of energy equal to $|\epsilon(\pi)|$. 
	Therefore, it would be energetically favorable to populate this fermionic mode. 
	However, the $\pi$-mode exists only in the even parity sector where the addition of a single excitation is forbidden by the parity constraint.
	Therefore to obtain a physical state in which the $\pi$-mode is populated we have to consider a state with two fermions in which the second lives in a different fermionic mode.
	But the addition of such a second fermion raises the energy by an amount that is greater than $\epsilon(\pi)$.  
	For this reason, the lowest energy state of the even sector in this region is still its Bogoliubov vacuum $\ket{\emptyset^+}$.
	In the odd sector, states with a single excitation are allowed, but all fermionic modes hold positive energy and, it is easy to check that each state that can be defined in this sector has an energy greater than the one associated to $\ket{\emptyset^+}$. Despite this, the lowest admissible states in this sector, those with one occupied mode with momentum closest to $\pi$ (exactly $\pi$ is not possible because of the quantization rule of this sector) have an energy gap closing as $1/N^2$ compared to $\ket{\emptyset^+}$.
	
	Therefore, the ground state of the whole Hamiltonian is still the Bogoliubov vacuum $\ket{\emptyset^+}$. 
	However, differently from the previous case, the ground state is no longer separated by a finite energy gap.
	In fact, due to the form of the dispersion relation, one can easily see that the energy gap closes quadratically in the thermodynamic limit and that there is an alternation between states with different parity.
	
	\item[\textbf{III}]
	$h<h^*=1-\gamma^2$: Here, the dispersion relation \eqref{eq:lambda} develops two symmetric minima at $q=\pm \tilde(q)$, with
\begin{equation}
\label{eq:ksdef}
    \tilde{q} =\arccos \left(\frac{h}{\gamma^2-1} \right).
\end{equation}
Note that the threshold parabola $h=1-\gamma^2$ differs from the usual circle $h^2 + \gamma^2 =1$ that encloses the region with oscillatory contributions to the correlation functions in the non-frustrated \textit{XY} chain.
In general, for finite size systems, $\tilde{q}$ is not an allowed lattice momentum of the system. 
Therefore let us define $\tilde{q}^+$ and $\tilde{q}^-$ as the momenta in the even and in the odd sectors closest 
to $\tilde{q}$. 
Since in the odd parity sector, there is no fermionic mode with a negative energy contribution, the lowest energy is associated to the states $\ket{\pm \tilde{q}^-}=b^\dagger_{\pm \tilde{q}^-} \ket{\emptyset^-} $.
On the contrary, in the even parity sector, 
the energy of the $\pm \tilde{q}^+$ modes is smaller than $1-h$. Thus, because the net contribution of the $\pi$ and the $\pm \tilde{q}^+$ modes provides am overall negative energy, the two lowest energy states in the even sector are $\ket{\pm \tilde{q}^+}=b^\dagger_{\pm \tilde{q}^+} b^\dagger_{\pi} \ket{\emptyset^+}$.
The four states, i.e. $\ket{\pm \tilde{q}^+}$ and $\ket{\pm \tilde{q}^-}$  have a lattice momentum respectively equal to $\pm\left(\pi + \tilde{q}^+\right)$ and $\pm \tilde{q}^-$~\cite{Maric2020_induced} and their associated energies are:
\begin{subequations}
\begin{eqnarray}
\label{low_energy_even}
E_e&=&\bra{\pm\tilde{q}^+}H\ket{\pm\tilde{q}^+}
\nonumber \\
&= &\Lambda(\tilde{q}^+)+\frac{\epsilon(\pi)}{2}-\frac{1}{2} \!\!\!\!\!\!\sum_{q\in \Gamma^+/\{\pi\}}\!\!\!\!\!\!\Lambda(q); \ \ \ \ \\
  \label{low_energy_odd}
 E_o&=&\bra{\pm\tilde{q}^-}H\ket{\pm\tilde{q}^-}
 \nonumber \\
 &=&\Lambda(\tilde{q}^-)-\frac{\epsilon(0)}{2}-\frac{1}{2} \!\!\!\!\!\!\sum_{q\in \Gamma^-/\{0\}}\!\!\!\!\!\!\Lambda(q). \ \ \  \ 
\end{eqnarray}
\end{subequations}
In each sector, the lowest energy state is separated from the other eigenstates by a gap that closes like $1/N^2$, with the lightest states having finite momenta close to $\tilde{q}$, as can be easily understood from the dispersion relation in eq.~\eqref{eq:lambda}. Thus, $\tilde{q}^\pm$ act as an effective Fermi momentum and one can think of the states spanning each ground state manifold as resulting from a shell-filling effect that keeps only one of the two Fermi points occupied.
To better understand the ground state structure, let us start with the assumption
that we choose the parameters of the Hamiltonian such that $\tilde{q}$ coincides with $ \tilde{q}^+$. 
Thus, the ground-state manifold is spanned
by the states $\ket{\pm\tilde{q}^+}$ and belongs to the even parity sector.
A generic change in the parameters $h$ and $\gamma$ will move $\tilde{q}$ away from $ \tilde{q}^+$ and bring it progressively closer to $ \tilde{q}^- $. At some point $E_o$ becomes smaller 
than $E_e$ and there will be a crossover between the states $\ket{\pm\tilde{q}^+}$ and $\ket{\pm\tilde{q}^-}$, with the ground-state manifold switching to the odd parity sector. 
Further moving the parameters in the same direction, eventually $\tilde{q}$ will come close to a different allowed momentum in the even sector. 
Hence the system will face a second crossover, and this process will continue until the parameters of the Hamiltonian exit from the chiral region $|h|<h^*$ and settle into the even parity fermionic vacuum as the ground state.
Increasing the dimension of the system, the distance between the different momenta becomes progressively smaller and hence the crossing become denser until each point in the region will be characterized by a crossover between two two-fold degenerate manifold belonging to two different parity sectors and having  different quantum numbers of the lattice momentum. 

In this portrayal, we have assumed that changes in the system parameters always change $\tilde{q}$. However, 
from eq.~\eqref{eq:ksdef}, we see that we can keep the minimum of the dispersion relation fixed by moving along the parabola $h=\mathrm{c} (1-\gamma^2)$, where $\mathrm{c}$ is a constant defined in the interval $[0,1]$. 
In chains of finite length, one can identify a strip around each parabola in which the ground state manifold remains constant and this strip becomes narrower as the chain length increases. In this way
the system undergoes a foliation of the ground state space that is made of as many manifolds as the number of sites in the system.
Let us stress once more that, any time a change of parameters changes $\tilde{q}^\pm$, hence crossing a strip, even for small systems the fidelity suddenly drops to zero, thus representing an extreme instance of orthogonality catastrophe \cite{Anderson1967-1, Anderson1967-2}.

{To address whether the chiral region represents a different thermodynamic phase or not, we look more closely at the energies.}
On one side, although the gap between the ground state and the first excited state closes as $ 1 / N^2$ in the whole frustrated phase for $|h|<1$, for $h^* <h <1$ the ground state is always represented by the fermionic vacuum state $ \ket{\emptyset^+} $, while in the chiral region the ground state manifold keeps changing its parity and momenta. Even increasing the chain length without moving $h$ and $\gamma$ can switch the ground state parity (see Fig.~\ref{fig:parity_gs}). Moreover, as can be appreciated from Fig.~\ref{fig:parity_gs}, the gap between the alternating ground states closes exponentially with the chain length, which means that in the thermodynamic limit the two manifold become effectively degenerate: crossing the line $h=h^*$ the ground state degeneracy thus grows from $1$ to $4$, which could indicate a first order quantum phase transition.
\begin{figure}[t]
\includegraphics[width=0.9\columnwidth]{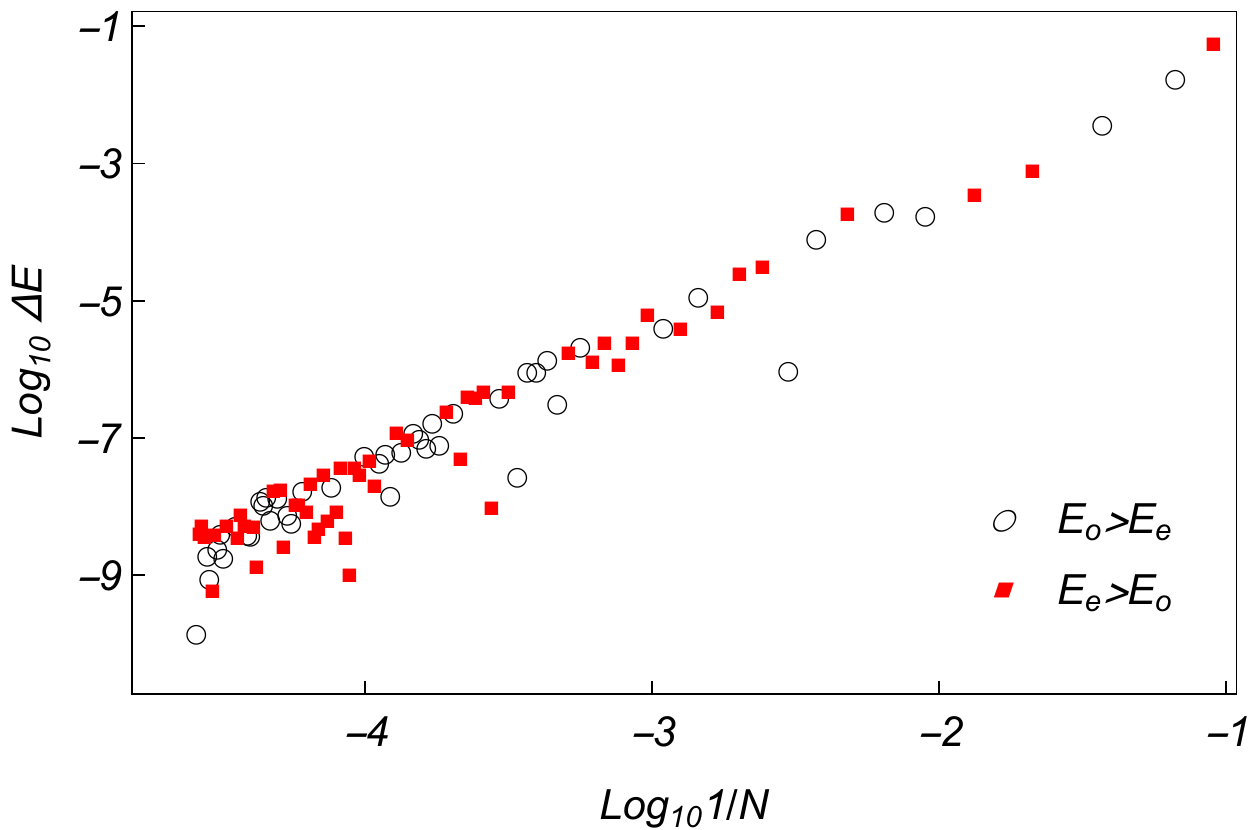}
\caption{Behavior of $\Delta E=|E_e-E_0|$, for $h=0.4$ and  $\gamma=0.6$ as function of the system length $N$. The red squares represent points in which the states in the even sector have energy greater than the ones in the odd sector, while the black circle signals the presence of a ground state manifold living in the even sector. The parity of the ground state manifold keeps alternating as more sites are added.} 
\label{fig:parity_gs}
\end{figure}
However, analyzing the free energy derivatives (which at zero temperature coincide with the ground state energy) we cannot detect any discontinuity that remains finite in the thermodynamic limit. For $h>h^*$
the ground state energy associated with the fermionic vacuum $\ket{\emptyset^+}$ is
\begin{eqnarray}
\label{low_energy_void}
  E_\emptyset &=\bra{\emptyset^+}H\ket{\emptyset^+}&=-\frac{\epsilon(\pi)}{2}-\frac{1}{2} \!\!\!\!\!\!\sum_{q\in \Gamma^+/\{\pi\}}\!\!\!\!\!\!\Lambda(q).
\end{eqnarray}
Immediately crossing into the chiral region, 
the ground state  energy is the odd state one in eq.~\eqref{low_energy_odd} with $\tilde{q}^-=\pi\left(1-\frac{1}{N}\right)$.
Starting from these two expressions for the energy below and above $h=h^*$ it is possible to study the behavior of the derivatives at any order at the two sides of this line.
Selecting a curve in the $(h,\gamma)$ space, parametrized by a parameter $\alpha$, which crosses the $h=h^*$ curve at $\alpha=0$, we compute
\begin{equation}
	\Delta(n)= \left. \frac{\partial^n E_o}{\partial \alpha^n} \right|_{\tilde{q}^-=\pi(1-1/N)}-\frac{\partial^n E_\emptyset}{\partial \alpha^n} ,
\end{equation}
which must be finite, in the thermodynamic limit, to signal a traditional phase transition.


In Fig.~\ref{fig:diff_derivative} we show an instance of the dependence of $\Delta(n)$ for $n$ running from $1$ to $3$ when we cross the line $h=h^*$ keeping $\gamma$ fixed.
As we can see below and above $h^*$, in the case of finite size systems, all the derivatives show a non-zero $\Delta(n)$, but these differences vanish proportionally to $1/N^2$.
This fact implies that in the thermodynamic limit all the derivatives are analytical and hence that if $h=h^*$ represents a quantum phase transition,  it has to be one akin to a BKT transition~\cite{BKT1,BKT2,BKT3}.
\end{itemize}

\section{Global state fidelity and fidelity susceptibility}

In order to evaluate the ground state fidelity in eq.~\eqref{eq:fidelity}, we observe that all ground-states can be written starting from the vacuum states $\ket{\emptyset^+}$ and $\ket{\emptyset^-}$. 
In terms of the Bogoliubov angles $\theta_q$, they can be formulated like
\begin{subequations}
	\label{eq:groundstate-a}
	\begin{eqnarray}
		\label{eq:groundstate-e-a}
		\ket{\emptyset^+}\!&\!=\!&\!\ket{0_\pi}\!\bigotimes_{q\in\Gamma_2^+}\!\left(\cos{\theta_q}\ket{0}_q\ket{0}_{-q}\!-\!\sin{\theta_q}\ket{1}_q\ket{1}_{-q} \right) \;\;\; \;\;\; \\
		\label{eq:groundstate-o-a}
		\ket{\emptyset^-}\!&\!=\!&\!\ket{0_0}\!\bigotimes_{q\in\Gamma_2^-}\!\left(\cos{\theta_q}\ket{0}_q\ket{0}_{-q}\!-\!\sin{\theta_q}\ket{1}_q\ket{1}_{-q} \right)\;\;\;\; \;\;
	\end{eqnarray}
\end{subequations}
where $\Gamma^+_2$ ($\Gamma^-_2$) is the subset of momenta $q \in \Gamma^+$ ( $q \in \Gamma^-$) that live in the interval $q\in(0,\pi)$.
From these expressions it is easy to obtain that, in the region $h>h^*$ where the ground state coincide with $\ket{\emptyset^+}$, the ground-state fidelity, becomes
\begin{equation}
	\label{eq:fidelity_0-a}
	\mathcal{F}=\prod_{q\in\Gamma^+_2}\cos(\theta'_q-\theta_q)\,,
\end{equation}
where $\theta_q$ ($\theta'_q$) are the Bogoliubov angles associated to the set of parameters $\overrightarrow{\lambda}=\{\gamma,h\}$
\mbox{$(\overrightarrow{\lambda}\!+\!d\overrightarrow{\lambda}=\{\gamma\!+\!d\gamma,h\!+\!dh\})$.}

The situation completely changes entering in the region $|h| <h^*$. 
In this case, the ground state is no more represented by the vacuum state in the even sector, but by states obtained by populating 1 or 2 fermionic levels, depending on the parity sector in which the ground-state manifold lives. 
In this situation, we have two different cases.
The first is when the two ground states are characterized by having either different numbers of fermions or fermions living in different modes.
These occur, in the thermodynamic limit, if the two sets of parameters are not on the same parabola $ h = \mathrm{c} (1 - \gamma^2) $ where $ \mathrm{c} $ is a constant obeying to the constraint $|\mathrm{c}|\le1$.
In this case, the overlap between the two states is reduced to the expectation value of either operators like $ b^\dagger_q b_{q'} $ with $ q \neq q' $, or of single creation (annihilation) operators on the vacuum states. 
But, from the expression of the vacuum states in eqs.~\eqref{eq:groundstate-a} it is easy to see that this expectation value is zero, and therefore also the ground-state fidelity vanishes.
On the contrary, in the case in which both ground states associated with the two sets of parameters are obtained from the vacuum states by exciting the same fermionic levels, i.e., for large $N$, if they are both on the same parabola $ h = \mathrm{c} (1 - \gamma^2) $, the ground-state fidelity does not cancel out identically. 
Hence, depending on the parity, in the even sector the ground-state fidelity becomes that in eqs.~(\ref{eq:fidelity_even},\ref{eq:fidelity_odd}).

Differently from all the other paths in the region $|h|<h^*$, along these parabolas it is possible to evaluate also the fidelity susceptibility that, by definition, is equal to the leading order of the expansion of the ground-state fidelity eq.~\eqref{eq:fidelsusc}.
The susceptibility $\chi$ is expected to be proportional to the system size in the non critical phase. In the case the ground-states are in the even sector it reads
\begin{equation}
	\label{eq:g-a}
	\chi =\!\!\!\!\!\!\!\!\!\sum_{q\in\Gamma^+/\{\tilde{q}^+\}}\!\!\!\!\!\left(\frac{\sin \theta_q (\mathrm{c}(1+\gamma^2) -\cos\theta_q)}{2(\gamma^2\sin^2\theta_q+(\cos\theta_q+\mathrm{c}(\gamma^2-1))^2)}\right)^2\!\!\!
\end{equation}
while in the odd sector the expression of the terms inside the sum is the same with the sum extending to $q\in\Gamma^-/\{\tilde{q}^-\}$. 
By introducing the normalized fidelity susceptibility $\tilde{\chi}\equiv \chi/N$, in the thermodynamic limit, we are able to obtain a result independent from the parity sector of the ground-state
\begin{eqnarray}
	\tilde{\chi}&=&\frac{1}{8\pi}\int_0^{\pi} dx \left[ \frac{\sin x \left(c(1+\gamma^2) -\cos x\right)}{[\mathrm{c}(\gamma^2-1) + \cos x]^2 +\gamma^2\sin^2 x}\right]^2 
\end{eqnarray}
This integral can be solved analytically using contour integration in the complex plane, upon changing variable to $z=e^{ix}$ obtaining eq.~\eqref{tildechi}.

\section{Majorana Correlation functions}
\label{ap:Majorana_correlations}

By knowing the ground-states in the different regions of parameter space, it is possible to evaluate the spin correlation functions following the approach described in detail in Ref.~\cite{Barouch1971}. 
It is based on the introduction of the Majorana fermionic operators $A_{i}$ and $B_{i}$ defined as
\begin{equation}
	\label{eq:Majorana-a}
	A_{i}\equiv \psi_{i}^{\dagger}+\psi_{i}; \quad B_{i}\equiv \imath \left(\psi_{i}-\psi_{i}^{\dagger} \right)\,,
\end{equation}
and the use of Wick's theorem~\cite{Wick1950}.
Indeed, each spin correlation function in which we are interested can be mapped, with the help of the Jordan-Wigner transformation in~\eqref{eq:JW-a} and the definition in~\eqref{eq:Majorana-a} in a string of Majorana operators on different spins. 
Then, with the help of Wick's theorem, the expectation value of such a string can be reduced to a Pfaffian in which each single element is the expectation value of a product of two different Majorana operators. 
Thereby each spin correlation function can be reduced to the evaluation of a  particular function of four kind of expectation values i.e. $\langle A_l A_j \rangle$, $\langle B_l B_j \rangle$, $\langle A_l B_j \rangle$ and $\langle A_j B_l  \rangle=-\langle  B_l A_j\rangle$ where $\langle \cdot \rangle$ stands for the expectation value on a specific ground-state and the indices $l$ and $j$ run over all sites of the system. 

For $h>1$ the ground state of the system has the form of eq.~\eqref{eq:groundstate-a} and it is easy to obtain for the Majorana correlation functions $\langle B_l A_j\rangle$ 
\begin{eqnarray}
	\label{eq:Majorana_1-a}
	\big\langle B_{l}A_{j} \big\rangle &= &\frac{\imath}{N}  \!\! \sum_{q\in\Gamma^+} \left[\sin 2\theta_{q} \sin\left( q r \right) + \cos 2\theta_{q} \cos\left( q r \right)  \right] \;\;\;\;\;\; 
\end{eqnarray}
where, for brevity, we have defined $r=j-l$, while $\langle A_l A_j\rangle=\langle B_l B_j\rangle=\delta_{l,j}$ .
For $h^*<h<1$ the ground state of the system still has the form of eq.~\eqref{eq:groundstate-a}, but Majorana correlation functions change. In fact, 
moving from values of the local field greater than 1 to less than 1, we have that while in the first case the energy associated with the momentum $q=\pi$ was positive, now it turns negative.
Hence, while $\langle A_l A_j\rangle$ and $\langle B_l B_j\rangle$ remain equal to $\delta_{l,j}$, the change in the sign of the energy of the fermionic mode induces a change in the sign of its contribution to the Majorana correlation functions $\langle A_l B_j\rangle$ that become
\begin{eqnarray}
	\label{eq:Majorana_2-a}
	\big\langle B_{j}A_{j+r} \big\rangle &= &\frac{2\imath(-1)^r}{N}+\frac{\imath}{N}   \sum_{q\in\Gamma^+} \sin 2\theta_{q} \sin\left( q r \right) \nonumber \\
	& & +\frac{\imath}{N}   \sum_{q\in\Gamma^+}  \cos 2\theta_{q} \cos\left( q r \right)  \;\;\;\;\;\; 
\end{eqnarray}

For $0\le h<h^*$ the situation becomes more complex since not only we have a dense series of crossovers between even and odd states, but also because each single ground-state manifold is two-fold degenerate even for finite $N$.
Let us consider the two cases, i.e. the manifold living in the even or the odd sector, separately. 
In the latter case, i.e. if the ground-state manifold falls in the odd-parity sector, all its elements can be written as 
\begin{equation}
	\ket{g^-}=\left(u b^\dagger_{\tilde{q}^-}+v b^\dagger_{-\tilde{q}^-}\right) \ket{\emptyset^-}
\end{equation}
where $u$ and $v$ are complex coefficients obeying the normalization conditions $|u|^2+|v|^2=1$.
Due to the presence of the fermions in the modes $\pm {\tilde{q}^-}$ we obtain that the Majorana correlation functions $\langle A_l A_j\rangle$ and $\langle B_l B_j\rangle$ are no more equal to zero when $l\neq j$, but become
\begin{equation}
	\label{eq:Majorana_3_1-a}
	\langle A_j A_{j+r}\rangle\!=\!\langle B_j B_{j+r}\rangle\!=\!\delta_{0,r}\!+\!\frac{2\imath}{N}(|v|^2\!-\!|u|^2)\sin\left(r \tilde{q}^-\right).
\end{equation} 
Moreover, also the correlation functions $\langle A_l B_j\rangle$ acquire a state dependent correction and become
\begin{eqnarray}
	\label{eq:Majorana_3_2-a}
	\big\langle B_{j}B_{j+r} \big\rangle \!&\!=\! &\! \frac{\imath}{N}   \sum_{q\in\Gamma^-} \left[\sin 2\theta_{q} \sin\left( q r \right) + \cos 2\theta_{q} \cos\left( q r \right)  \right] \nonumber \\ 
	& & -\frac{2\imath}{N} \big[ \sin 2\theta_{\tilde{q}^-} \sin\left( r \tilde{q}^- \right) 
	+  \cos 2\theta_{\tilde{q}^-} \cos\left(r \tilde{q}^- \right)\big] \nonumber \\
	\! &\! \!&\!  + 4|u v^*| \cos(\tilde{q}^-(r+2j)+\alpha ) 
\end{eqnarray}
where $\alpha$ is the phase of the complex number $uv^*$.

On the other hand, in the even sector of the parity the general ground state can be written as
\begin{equation}
	\ket{g^+}= b_\pi^\dagger \left(u b^\dagger_{\tilde{q}^+}+v b^\dagger_{-\tilde{q}^+}\right) \ket{\emptyset^+}
\end{equation}
Similarly to the odd case we recover
\begin{equation}
	\label{eq:Majorana_3_3-a}
	\langle A_j A_{j+r}\rangle\!=\!\langle B_j B_{j+r}\rangle\!=\!\delta_{0,r}\!+\!\frac{2\imath}{N}(|v|^2\!-\!|u|^2)\sin\left(r \tilde{q}^+\right).
\end{equation} 
and
\begin{eqnarray}
	\label{eq:Majorana_3_4-a}
	\big\langle B_{j}A_{j+r} \big\rangle \!&\!=\! &\! \frac{\imath}{N}   \sum_{q\in\Gamma^+} \left[\sin 2\theta_{q} \sin\left( q r \right) + \cos 2\theta_{q} \cos\left( q r \right)  \right] \nonumber \\ 
	& & -\frac{2\imath}{N} \big[ \sin 2\theta_{\tilde{q}^+} \sin\left( r \tilde{q}^+ \right) 
	+  \cos 2\theta_{\tilde{q}^+} \cos\left(r \tilde{q}^+ \right)\big] \nonumber \\
	\! &\! \!&\!  + 4|u v^*| \cos(\tilde{q}^+(r+2j)+\alpha ) 
\end{eqnarray}

\end{document}